\documentclass[aps,prl,reprint,citeautoscript,twocolumn,superscriptaddress,longbibliography]{revtex4-2}
\usepackage[english]{babel}
\usepackage{amsmath,amssymb}
\usepackage{multirow}
\usepackage{bm}
\usepackage{mathtools}
\usepackage{dsfont}
\usepackage{amsfonts}
\usepackage{amsthm}
\usepackage[normalem]{ulem}
\usepackage{url}
\usepackage{wasysym}
\usepackage{bbm}
\usepackage{quantikz}
\usepackage{float} 
\usepackage{xcolor}
\usepackage{bibunits}
\usepackage{etoolbox}

\usepackage{MnSymbol}
\usepackage{times}
\makeatletter
\renewcommand{\@bibunitname}{\jobname-bu\the\@bibunitauxcnt}
\makeatother

\definecolor{tum}{HTML}{0065bd}
\definecolor{arxivcite}{HTML}{004D6B}
\definecolor{arxivlink}{HTML}{8c1515}

\usepackage[colorlinks=true,linkcolor=arxivlink,citecolor=arxivcite,urlcolor=arxivlink,hypertexnames=false]{hyperref}
\usepackage{orcidlink}

\newcommand{\SMAttachedAddToContents}[2]{%
    \ifstrequal{#1}{toc}{\SMaddtocontents{smtoc}{#2}}{\SMaddtocontents{#1}{#2}}%
}
\makeatletter
\AtBeginDocument{%
    \let\SMsectioncntformat\@sectioncntformat
    \let\SMhangfromsection\@hangfrom@section

}
\makeatother

\def\ba#1\ea{\begin{align}#1\end{align}}
\def\bg#1\eg{\begin{gather}#1\end{gather}}
\def\bpm{\begin{pmatrix}}
\def\epm{\end{pmatrix}}

\newcommand{\iu}{\mathrm{i}}

\mathchardef\mhyphen="2D

\newcommand{\ourtitle}{Fractionalized Chiral Edge Dynamics from Initial Product States}
\allowdisplaybreaks

\begin{document}
\title{\textbf{\ourtitle}}
%%%%%%%%%%%%

%%%%%%%%%%%%
\author{Shi Feng\,\orcidlink{0000-0002-0184-7049}}
\email{shi.feng@tum.de}
\affiliation{Technical University of Munich, TUM School of Natural Sciences, Physics Department, 85748 Garching, Germany}
\affiliation{Munich Center for Quantum Science and Technology (MCQST), Schellingstr. 4, 80799 M{\"u}nchen, Germany}
\author{Willian M. H. Natori\,\orcidlink{0000-0002-0740-2956}}
\affiliation{Gleb Wataghin Institute of Physics, University of Campinas, Campinas,
São Paulo 13083-950, Brazil}
\author{Michael Knap\,\orcidlink{0000-0002-7093-9502}}
\affiliation{Technical University of Munich, TUM School of Natural Sciences, Physics Department, 85748 Garching, Germany}
\affiliation{Munich Center for Quantum Science and Technology (MCQST), Schellingstr. 4, 80799 M{\"u}nchen, Germany}
\author{Johannes Knolle\,\orcidlink{0000-0002-0956-2419}}
\affiliation{Technical University of Munich, TUM School of Natural Sciences, Physics Department, 85748 Garching, Germany}
\affiliation{Munich Center for Quantum Science and Technology (MCQST), Schellingstr. 4, 80799 M{\"u}nchen, Germany}
\affiliation{Blackett Laboratory, Imperial College London, London SW7 2AZ, United Kingdom}

%%%%%%%%%%%%

%%%%%%%%%%%%
\begin{abstract}
Quantum simulation of fractionalized chiral edge modes is often assumed to require the preparation of a long-range-entangled topological state. Here we show that the dynamics of fractionalized chiral edge modes in Kitaev-type spin liquids can instead be sharply resolved from simple high-energy product states prepared by only single-qubit rotations. These states select the global chirality by fixing a subset of fluxes which break time reversal symmetry on plaquettes with odd number of sites, while at the same time leaving the extensive set of even plaquettes' fluxes unresolved. The latter flux defects  bind  Majorana zero modes akin to Aharonov–Bohm--cages whose compact support prevents hybridization, preserving the fractionalized band topology despite strong flux disorder. We show that spin correlations of quench dynamics intrinsically average over these unresolved  flux sectors, that dynamically dephase nonuniversal oscillations and reveal sharp chiral recurrences and direct signatures of chiral propagation. We demonstrate this mechanism in the Yao–Kivelson model and construct a digital protocol that preserves the Aharonov–Bohm--caged Majorana zero modes despite finite Trotter errors. Our mechanism extends to other variants of Kitaev models, offering a route to observing fractionalized chiral dynamics with simple state preparation across quantum-simulation platforms.
\end{abstract}
%%%%%%%%%%%%

%%%%%%%%%%%%
\maketitle 
% \tableofcontents

\emph{ \color{tum}{Introduction.--}}
Recent advances in synthetic quantum platforms have fueled interest in simulating topological quantum matter~\cite{Satzinger2021,Semeghini21,Pollmann26}. Quantum spin liquids (QSLs) exhibiting topological order and fractionalization \cite{savary2017quantum,Senthil20} are particularly compelling targets that have been intensely studied in this context. 
Their simulation complements long-standing efforts to observe these exotic phenomena in solid-state materials~\cite{knolle2019field}. Recently, Kitaev QSLs~\cite{Kitaev2006} have been searched for in spin-orbit coupled magnetic materials~\cite{hermanns2018physics,TREBST20221,Matsuda25,Feng2026} and their topology probed on  synthetic platforms, including superconducting processors~\cite{Will2025,Park25}, Rydberg-atom arrays~\cite{Evered2025}, and trapped-ion processors~\cite{ali2025}.

Topology in QSLs can enter at two intertwined but distinct levels: topological order encoded in the non-trivial many-body correlations of a (ground) state, and the emergent band topology of fractionalized particles. The exactly soluble Kitaev honeycomb model~\cite{Kitaev2006} connects these aspects explicitly: its emergent $\mathbb Z_2$ gauge field contributes to the ground state's topological entanglement entropy~\cite{Hong10}, and sets the flux background in which fractionalized Majorana fermions form topological Chern bands. 
The former concerns the long-range entanglement in a many-body state; while the latter characterizes the band topology of matter excitations.
Such distinction between the two connotations of topology suggests that their effective probes may differ, allowing each aspect to be targeted separately. This motivates the question whether the emergent band topology of fractionalized Majorana fermions can be resolved without preparing a topologically ordered state?

Two obstacles arise here. (I) A short-range-entangled high-energy spin state generally is a superposition of many \emph{disordered} flux sectors, even when the state is translation-invariant. This parallels the emergent disorder in disorder-free localization, where an infinite-temperature initial state samples different configurations of gauge super-selection sectors~\cite{Knolle17,Google26,Halimeh26}. However, extensive flux defects can then hybridize their bound Majorana zero modes (MZMs) into an extended low-energy continuum, closing the mobility gap needed for chiral band topology~\cite{Huse2012}. Moreover, generic flux disorder can lead to slow  glassy dynamics of the fermionic excitations~\cite{ZhuHeyl2021,Feng2025}. (II) Such an initial state lies far from equilibrium, so the relevance of low-energy universal chiral dynamics is \emph{a priori} unclear.
These obstacles are the reason why existing proposals for quantum simulations of the Kitaev QSL have relied on the preparation of a homogeneous flux sector \cite{JinThermalization2023,Marin23} or the full topologically ordered ground state combining flux-error correction with variational optimization of the Majorana fermion vacuum~\cite{Evered2025,ali2025}.

\begin{figure*}
    \centering
    \includegraphics[width=0.98\linewidth]{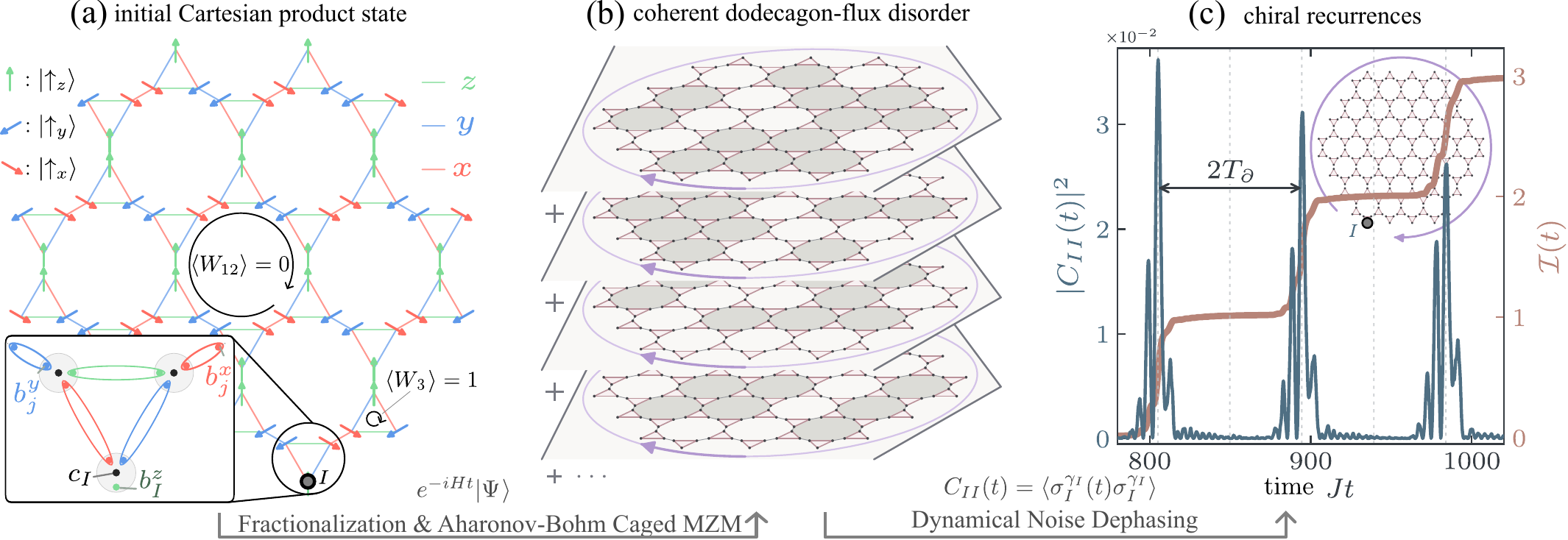}
    \caption{\textbf{Fractionalized chiral dynamics from a spin product state.}
    (a) Bond colors denote the $x,y,z$ spin exchanges in the Yao-Kivelson model. Arrows show the Cartesian spin product state that fixes every triangular flux to $W_3=1$, while leaving dodecagon fluxes unresolved. The inset illustrates the fractionalization of spins into matter and gauge Majorana fermions; the $\mathbb Z_2$ fields $\hat{u}_{ij} = ib_i^\alpha b_j^\alpha$ are denoted by ovals. 
    (b) In the fractionalized basis, this state is a coherent superposition of dodecagon flux sectors (layers), with flipped fluxes shaded gray and the fixed $W_3=1$ triangular fluxes shaded pink. Their Aharonov--Bohm--caged compact MZMs do not hybridize, preserving the chiral boundary channel (purple) across the ensemble.
    (c) Simulated chiral recurrence by the setup shown in the inset. Flux-sector average from the correlator $C_{II} = \langle \sigma_I^{\gamma_I}(t) \sigma_I^{\gamma_I}\rangle$ dephases non-universal oscillations, revealing sharp boundary-spin recurrences with approximately integer steps.}
    \label{fig:blurb}
\end{figure*}

In this Letter, we present a mechanism that overcomes both obstacles and dynamically reveals fractionalized band topology of Kitaev-type chiral QSLs from a \emph{spin product state}, readily prepared by single-qubit rotations as illustrated in Fig.~\ref{fig:blurb}(a), without optimizing a flux configuration or fermion vacuum. Specifically, (I) is overcome by a suitably chosen product state that restricts the emergent flux disorder to defects whose bound MZMs are Aharonov--Bohm caged \cite{Vidal98,Faini01}, meaning they have strictly local support. As a result, the mobility gap and chiral band topology are preserved. (II) is overcome because quench spin correlations intrinsically average over disordered flux sectors, dynamically dephasing non-universal oscillations and revealing universal chiral edge dynamics [Fig.~\ref{fig:blurb}(b,c)]. We demonstrate this mechanism in the Yao-Kivelson chiral QSL model~\cite{YaoKivelson2007}, though it applies broadly to many other variants of decorated Kitaev models. We discuss its robustness and construct a digital circuit that stroboscopically stabilizes the Aharonov--Bohm--caged MZMs and chiral dynamics in presence of Trotter errors. Concrete proposals for different quantum processors and further details are presented in the Supplemental Material (SM) \cite{supp}.

\emph{\color{tum}Decorated Kitaev model.--}
We consider triangle-decorated Kitaev-type spin-$\frac{1}{2}$ models, with exchange $J_\triangle$ on bonds within triangles and $J_\parallel$ on bonds connecting them:
\begin{equation}
    H=-J_\triangle \sum_{\langle ij\rangle_\alpha \in \triangle}
    \sigma_i^\alpha\sigma_j^\alpha
    - J_\parallel \sum_{\langle ij\rangle_\alpha \in \parallel}
    \sigma_i^\alpha\sigma_j^\alpha
    \label{eq:YK_spin_H}
\end{equation}
Here $\langle ij\rangle_\alpha$ denotes a nearest-neighbor bond of Kitaev coloring $\alpha=x,y,z$. Such lattices can be constructed by replacing each vertex of any trivalent parent lattice with a triangle.
The Yao-Kivelson model, composed of triangles and dodecagons~\cite{YaoKivelson2007}, provides the simplest example, see the tri-colored lattice shown in Fig.~\ref{fig:blurb}(a).
For concreteness, we focus on the chiral phase at $J_\parallel=J_\triangle=J$ where the chiral QSL ground state spontaneously breaks time-reversal symmetry. 
The exact solution represents each spin by Majorana operators: $\sigma_j^\alpha=\iu b_j^\alpha c_j$, subjected to the constraint $D_j=b_j^x b_j^y b_j^z c_j=1$ \cite{Kitaev2006}. Each oriented $\alpha$ bond carries a conserved $\mathbb Z_2$ gauge operator: $\hat u_{ij}=\iu b_i^\alpha b_j^\alpha$, shown in the inset of Fig.~\ref{fig:blurb}(a). Its gauge invariance is encoded in the plaquette fluxes,
$\hat W_{n(p)}=\prod_{j=1}^{n(p)}\sigma_j^{\gamma_j}
=-\prod_{j=1}^{n(p)}\hat u_{j,j+1}$,
with sites ordered around the $n(p)$-sided plaquette $p$ and $\gamma_j$ the color absent from its two bonds meeting at $j$. For a specified $\mathbb{Z}_2$ gauge $u$, the matter Majoranas are governed by
$H(u)=\frac{\iu}{4}\sum_{ij}A_{ij}(u)c_ic_j$,
with the antisymmetric hopping $A_{ij}(u)=2Ju_{ij}$ defined on each bond.
The two time-reversed ground-state flux sectors, labeled by $\tau = \pm 1$, have uniform triangular flux $W_3=\tau$ and dodecagon flux $W_{12}=+1$. Each ground state realizes a chiral spin liquid with Majorana Chern number $|C|=1$ as in the time-reversal-broken Kitaev honeycomb model \cite{Kitaev2006}, supporting chiral edge modes and flux-bound MZMs. Next, we discuss the initial product state for this model and the mechanism by which fractionalized chiral dynamics can be revealed.

\begin{figure*}[t]
    \centering
    \includegraphics[width=0.95\linewidth]{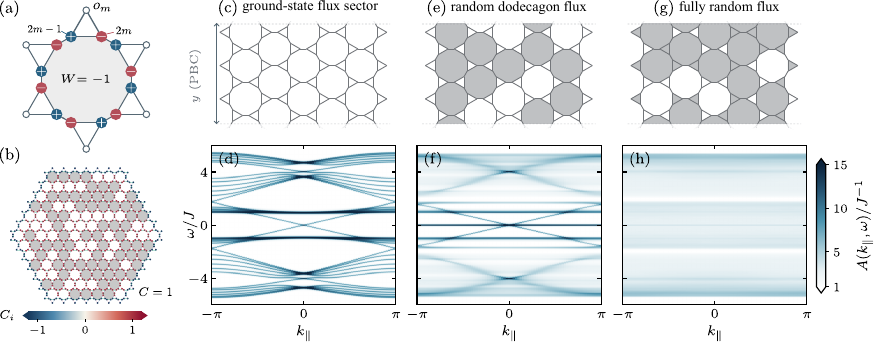}
\caption{
\textbf{Aharonov--Bohm--caged MZM, topological marker, and Majorana spectra.}
(a) A compact Aharonov--Bohm--caged MZM bound to a flipped dodecagon, $W_{12}=-1$. Blue and red sites carry amplitudes $\pm1/\sqrt{12}$; open sites and the exterior have zero amplitude.
(b) Local Chern marker under dodecagon flux disorder, with bulk average $C=1$.
(c,e,g) Unwrapped Yao--Kivelson cylinders illustrating the ground-state flux sector, a representative of dodecagon-flux disorder, and that of all-plaquette flux disorder, respectively. Gray plaquettes denote fluxes flipped relative to the ground-state sector.
(d,f,h) Corresponding average-momentum-resolved spectral functions $A(k_\parallel,\omega)$ on an $L_x=6$, $L_y=80$ cylinder. Dispersing edge branches survive in the clean (d) and dodecagon-disordered (f) sectors, with compact zero modes producing flat spectral weight at $\omega=0$ in (f). All-plaquette disorder (h) fills the spectral gap and removes the resolved edge branches. Disordered spectra average 1000 samples with independent flux-flip probability $p=1/2$.
}
    \label{fig:cmzm}
\end{figure*}

\emph{\color{tum}Cartesian Product State.--}
Consider the high-energy initial product state illustrated by the colored arrows in Fig.~\ref{fig:blurb}(a):
\begin{equation} \label{eq:state}
\lvert\Psi\rangle=\bigotimes_\triangle\lvert\psi_\triangle\rangle,
    \,\, \,
    \lvert\psi_\triangle\rangle
    =\lvert\uparrow_{\gamma_j}\rangle_j
    \otimes\lvert\uparrow_{\gamma_k}~\rangle_k
    \otimes\lvert\uparrow_{\gamma_l}~\rangle_l
\end{equation}
where the sites $j,k,l \in \triangle$ and $\lvert\uparrow_{\gamma_j}\rangle_j$ is polarized along the axis whose color is absent from the two intra-triangle bonds meeting at $j$. 
The key observation is that each triangle satisfies
$W_3\lvert\psi_\triangle\rangle=\lvert\psi_\triangle\rangle$,
fixing a common chirality throughout $\lvert\Psi\rangle$. In contrast, every dodecagon remains flux-unresolved by
$\langle\Psi|W_{12}|\Psi\rangle=0$, i.e. $\lvert\Psi\rangle$ is a superposition of all possible dodecagon flux sectors [Fig.~\ref{fig:blurb}(b)]. 
Note that the time-reversed state selects the opposite chirality and the same construction applies to other triangle-decorated trivalent lattices. We emphasize that it is not the unique state that chooses such a flux pattern; it is, however, among the simplest candidate states that can be prepared using single-qubit rotations. 

%In Kitaev's parton representation \cite{Kitaev2006}, it becomes a coherent superposition of dodecagon flux configurations and their conditional matter states [Fig.~\ref{fig:blurb}(b)].
For an open boundary, the parton representation also includes the dangling gauge Majorana modes at two-coordinated boundary sites. Pairing these into fermions defines a boundary occupation basis $\lvert\beta\rangle$.
Writing $\lvert\widetilde\Psi\rangle$ for the normalized embedding of the spin state in the parton Hilbert space~\cite{Feldmeier2020}, we obtain
\begin{equation}
 \begin{aligned}
 \lvert\widetilde\Psi\rangle
 =\sum_u^\prime\lvert u\rangle\otimes\lvert\chi_u\rangle,\,\,\,
 \lvert\chi_u\rangle
 \equiv\sum_\beta
 \sqrt{p_{u\beta}}\,e^{\iu\theta_{u\beta}}\,
 \lvert\varphi_{u\beta}\rangle\otimes
 \lvert\beta\rangle,
 \end{aligned}
 \label{eq:resolve}
\end{equation}
where $\lvert\chi_u\rangle$ is the joint state of dangling gauge fermions [e.g. $b_I^z$ in Fig.~\ref{fig:blurb}(a)'s inset] and matter Majoranas conditioned on $u$, with $\langle\chi_u|\chi_u\rangle=p_u=\sum_\beta p_{u\beta}$ the distribution of the relevant gauge field and $\sum_u^\prime p_u=1$, and $\theta_{u\beta}$ is an associated phase.
The primed sum restricts every triangular flux to $+1$, while allowing both dodecagon fluxes $\pm1$ with uniform weights.

Since the gauge field and dangling gauge Majoranas are conserved, the spin dynamics initialized with the Cartesian product state is governed by the coherent superposition of matter fermion in $\lvert\chi_{u}\rangle$ evolving under their own conditional Hamiltonians $H(u)$ with fixed triangular fluxes and random dodecagon fluxes, akin to the non-equilibrium dynamics in the context of disorder-free localization \cite{Knolle17,Google26}. The band topology of these Hamiltonians $\{H(u)\}$ determines which sectors in the sum of Eq.~\eqref{eq:resolve} support chiral edge dynamics.

Disorder in $H(u)$ generally destroys Majorana band topology: flux defects bind MZM with finite localization length $\xi$, allowing their wavefunctions to overlap and hybridize. This  can generate an extended low-energy continuum that closes the mobility gap and destroys topological bands of Majorana fermions, resulting in a weakly anti-localized Majorana metal \cite{Huse2012,Fulga20,penghao24,wang2024,Feng2025,Feng2026}, see also SM \cite{supp}. However, as will be detailed in the next section, the Cartesian product state Eq.~\eqref{eq:state} on the decorated Kitaev lattice prevents this hybridization.

\emph{\color{tum}Aharonov--Bohm--caged MZM.--}
For Majorana band topology to survive flux-bound MZMs must avoid hybridization, which turns out to be possible for selected flux disorder configurations leading to the MZM's localization length being confined to $\xi = 0$.  
This is achieved by an Aharonov--Bohm cage \cite{Vidal98,Faini01} in the conditional Majorana Hamiltonian $H(u)$. As illustrated in Fig.~\ref{fig:cmzm}(a): in a triangle-decorated lattice, the Majorana amplitudes on the bonds leaving an even-sided plaquette (here a dodecagon) cancel exactly, confining the mode to a finite cage with strictly compact support. 
Such caging is absent in the Kitaev honeycomb model~\cite{supp}. In the Yao-Kivelson model at $J_\parallel=J_\triangle$, a dodecagon supports a caged MZM with $\xi=0$ if and only if
\begin{equation} \label{eq:cage}
    W_{12}=-1, \;\; W_3=\tau\;\; {\rm for}\;\;\tau=\pm1,
\end{equation}
where $W_3=\tau$ is required on the six triangles sharing an edge with that dodecagon. This local criterion extends to all even-sided plaquettes in the triangle-decorated trivalent models of Eq.~\eqref{eq:YK_spin_H} at equal couplings, as proved in Appendix~A. 
These compact modes remain exact zero-energy states without hybridizing \cite{Tim26}, preserving the gap despite extensive even-flux disorder. The surviving topology is reflected in the quantized Chern number $C=1$ computed from the local topological marker \cite{Bianco11,supp}, see Fig.~\ref{fig:cmzm}(b).

Consistently, the dodecagon-disordered Majorana spectrum resolved in average momenta [Fig.~\ref{fig:cmzm}(e,f)] retains the sharp chiral edge modes of the clean system [Fig.~\ref{fig:cmzm}(c,d)], alongside a persistent mobility gap and the flat zero-energy weight from the caged MZMs, even though its disorder-sensitive high-energy modes are averaged to a continuum, see SM \cite{supp} for details. In contrast, flipping a triangular flux from a ground state introduces a MZM with larger spatial support, making hybridization possible with dense flux defects. The all-plaquette-disordered spectrum consequently loses its resolved chiral modes due to the flipped triangular chirality [Fig.~\ref{fig:cmzm}(g,h)]. Note that while these calculations give sharp evidence of the surviving band topology of Majorana fermions, they are based on specified gauge fields and are not gauge-invariant observables. We will discuss gauge-invariant spin dynamics in the next section. 

We note that the exact caging and thus persistent Majorana topology require fine-tuning: generic perturbations can spoil destructive interference and induce zero-mode hybridization, see SM \cite{supp}. Nevertheless, weak violations of compactness can still permit chiral dynamics at finite sizes and timescales. For digital evolution, Appendix~B further addresses this through a zero-mode-preserving Trotterization.

\emph{\color{tum}Chiral Recurrences.--}
We probe the dynamical spin correlation on a boundary spin $I$:
\begin{equation} \label{eq:cii}
    C_{II}(t) = \langle\Psi|\sigma_I^{\gamma_I}(t)\sigma_I^{\gamma_I}|\Psi\rangle
\end{equation}
where the axis $\gamma_I$ is associated with its absent third bond. Because $|\Psi\rangle$ is an eigenstate of $\sigma_I^{\gamma_I}$, it is also equivalent to a single-spin quench probe $\langle\Psi|\sigma_I^{\gamma_I}(t)|\Psi\rangle$.
The selected boundary-spin operator $\sigma_I^{\gamma_I}$ is $\mathbb{Z}_2$ gauge-invariant, and its associated dangling gauge Majorana $b_I^{\gamma_I}$ is conserved [inset of Fig.~\ref{fig:blurb}(a)]. 
Using the orthogonality between gauge fields~\cite{Baskaran2007} and the Gaussian propagator of fermions, Eq.~\eqref{eq:cii} and \eqref{eq:resolve} give
\begin{equation} \label{eq:ciia}
 C_{II}(t)=\sum_u^\prime\sum_r [e^{A(u)t}]_{Ir}
 \langle\chi_u|c_rc_I|\chi_u\rangle.
\end{equation}
Formally, Eq.~\eqref{eq:ciia} contains non-local Majorana propagators $[e^{A(u)t}]_{Ir}$ beyond the local return amplitude $[e^{A(u)t}]_{II}$, weighted by the initial correlations $\langle\chi_u|c_rc_I|\chi_u\rangle$. The spectral contributions of the former involve additional spatial complex phases that can cancel across flux sectors, and will be suppressed in the thermodynamic limit. Remarkably, the Cartesian product state eliminates these off-site contributions \emph{exactly} in every sector by the identity 
\footnote{The initial state is an eigenstate of $\sigma_I^{\gamma_I}$ satisfying $\sigma_I^{\gamma_I} \lvert \widetilde\Psi \rangle - \lvert \widetilde \Psi \rangle = 0$. 
Since the edge spin preserves every gauge link, substituting Eq.~\eqref{eq:resolve} gives $\sum_u^\prime|u\rangle\otimes(\sigma_I^{\gamma_I}|\chi_u\rangle-|\chi_u\rangle)=0$. Using $\langle v|u\rangle=\delta_{vu}$ gives $\sigma_I^{\gamma_I}|\chi_v\rangle=|\chi_v\rangle$. 
The bilinear $\sigma_I^{\gamma_I}=\iu b_I^{\gamma_I}c_I$ anticommutes with $c_I$ and commutes with $c_r$ for $r\neq I$. Thus $\langle\chi_u|c_rc_I|\chi_u\rangle=\allowbreak\langle\chi_u|\sigma_I^{\gamma_I}c_rc_I\sigma_I^{\gamma_I}|\chi_u\rangle=\allowbreak-\langle\chi_u|c_rc_I|\chi_u\rangle$, forcing $\langle\chi_u|c_rc_I|\chi_u\rangle=0$. For $r=I$, $c_I^2=1$ gives $\langle\chi_u|c_I^2|\chi_u\rangle=p_u$.}: $\langle\chi_u|c_rc_I|\chi_u\rangle=p_u\delta_{rI}$.
With Eq.~\eqref{eq:ciia}, the quenched spin correlator takes the exact and compact form:
\begin{equation}
 C_{II}(t)= \sum_u^\prime p_u[e^{A(u)t}]_{II}, 
 \label{eq:averaged_return}
\end{equation}
which averages Majorana return amplitudes over the chiral flux ensemble selected by the Cartesian product state. Remarkably, although the initial state has high energy density and flux-sector-dependent non-equilibrium oscillations, upon averaging, they dephase automatically within a transient timescale of $1/{\rm bandwidth}$, which suppresses the non-universal noise and exposes the chiral propagation common to all dodecagon-disordered sectors, see also SM \cite{supp}.

\begin{figure}[b]
 \centering
 \includegraphics[width=\linewidth]{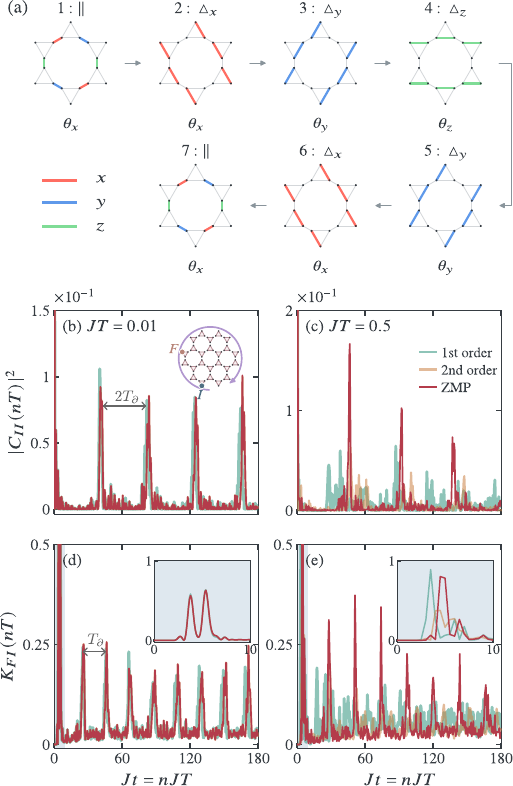}
 \caption{\textbf{Digital evolutions for on-site spin dynamics and OTOC.} (a) Illustration of the seven-layer ZMP cycle. The labels $\triangle_x,\triangle_y,\triangle_z$ select bonds within triangles. The $\parallel$ layers act on inter-triangle bonds, with angles from Eq.~\eqref{eq:app_digital_angles1}-\eqref{eq:app_digital_angles3}. (b,c) Chiral recurrences and (d,e) directional OTOCs with different Trotterizations, simulated on the 72-site lattice shown in the inset of (b), for different time steps $JT=0.01$ (b,d) and $0.5$ (c,e). Matching gray shading links the early-time windows within $0\leq Jt\leq10$ in (d,e) to their insets, which resolve the first arrival of chiral OTOCs. }
 \label{fig:digital_protocol}
\end{figure}

Numerical results using Eq.~\eqref{eq:averaged_return} are presented in Fig.~\ref{fig:blurb}(c), showing sharp recurrences separated by $2T_\partial$, where $T_\partial$ is the time to traverse the boundary once. The doubled periodicity is an interference effect: after an odd number of windings, enclosing an even or odd number of flipped dodecagon fluxes in the bulk contribute a $0$ or $\pi$ phase, which cancel on averaging; whereas recurrences from even windings are not affected. 
These recurrences are extremely sharp at early times [see also Fig.~\ref{fig:digital_protocol}(b,c), Appendix~B], while the initial chiral wave packet progressively delocalizes along the boundary, lowering and broadening its recurrences. Nevertheless, they remain well resolvable even at intermediate times before the wave packet spreads over the full perimeter. In this regime, non-universal oscillations have largely dephased and the low-energy chiral edge dynamics dominates. 
To quantify the intermediate- to late-time chiral dynamics, we obtain $\mathcal I(t)$ by integrating $|C_{II}(t)|^2$ from a time before a chosen reference recurrence, normalized so that the complete reference recurrence contributes one unit. Subsequent peaks broaden and weaken but contribute almost equal integrated weights, see Fig.~\ref{fig:blurb}(c).

%Edge-mode recurrences alone, however, cannot establish chirality since nonchiral edge modes can also recur in the same way. We therefore resolve the propagation direction via  spatially separated probes, as detailed in the next section. 

\begin{figure}[b]
 \centering
 \includegraphics[width=.9\linewidth]{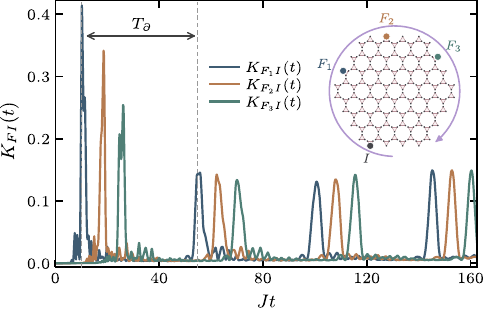}
 \caption{\textbf{Directional boundary OTOC from the Cartesian product state.} The inset shows the simulated Yao-Kivelson setting, with fixed triangle fluxes, unresolved dodecagon fluxes, and gauge-invariant spin probes at $I,F_1,F_2,F_3$. The operators lie six, ten, and fourteen boundary steps apart. The colored curves show $K_{F_aI}(t)$, averaged over the same 1000 independent even-flux configurations with flip probability $p=1/2$. }
 \label{fig:otoc_directional}
\end{figure}

\emph{ \color{tum}{Directional OTOC.--}}
A natural way to probe the chirality of the edge propagation would correlate spins at distinct boundary sites $I$ and $F$. However, being an eigenstate of the selected edge-spin operator, the off-site correlator on $\lvert \Psi \rangle$ is reduced to $C_{FI}(t)=\langle\Psi|\sigma_F^{\gamma_F}(t)\sigma_I^{\gamma_I}|\Psi\rangle=\langle\Psi|\sigma_F^{\gamma_F}(t)|\Psi\rangle$. This reproduces the same local return as studied in the previous section, and thus cannot resolve the propagation direction. We point out it is indeed possible to use $C_{FI}(t)$ as an analytically tractable probe of chirality by using different initial states and fractal lattices, which is beyond the scope of this work and will be discussed elsewhere \cite{companion_work}.

We here instead use the out-of-time-ordered correlator (OTOC) \cite{Swingle2016,google_otoc,XuSwingle2024} between boundary-spin commutators as a probe of directional operator spreading:
\begin{equation}
 K_{FI}(t)\equiv\frac{1}{2}
 \bigl\|[\sigma_F^{\gamma_F}(t),\sigma_I^{\gamma_I}]|\Psi\rangle\bigr\|^2,~~ I \neq F
 \label{eq:otoc_spin}
\end{equation}
Using $\sigma_F^{\gamma_F}(t)=\iu b_F^{\gamma_F}c_F^{(u)}(t)$, and the anticommutation between $c_F^{(u)}(t)$ and the dangling gauge Majorana fermions, one readily obtains $[\sigma_F^{\gamma_F}(t),\sigma_I^{\gamma_I}]=\allowbreak b_F^{\gamma_F}b_I^{\gamma_I}\{c_F^{(u)}(t),c_I\}$, giving the exact parton representation of edge-spin OTOC:
\begin{equation}
 K_{FI}(t)=2\sum_u^\prime p_u\bigl|[e^{A(u)t}]_{FI}\bigr|^2,
 \label{eq:otoc_majorana}
\end{equation}
which averages the square amplitudes of the Majorana propagators. These nonnegative real-valued summands do not dephase completely between flux sectors as the amplitudes in Eq.~\eqref{eq:averaged_return} do. Nevertheless, the chiral modes are insensitive to dodecagon-flux disorder [Fig.~\ref{fig:cmzm}(e,f)], and add constructively across flux sectors upon averaging, while non-universal oscillations vary between sectors and average to a smooth background that is much weaker than the peaks of winding chiral modes at intermediate timescales, see Fig.~\ref{fig:otoc_directional}. Squaring the propagation amplitudes before averaging in Eq.~\eqref{eq:otoc_majorana} removes the $\pi$-phase distinction between even and odd numbers of enclosed dodecagon fluxes and gives a recurrence time of $T_\partial$. 
%Note, the contrasting $2T_\partial$ period of $C_{II}(t)$ and the $T_\partial$ period of OTOC provide a clear signature for the interference with $\mathbb Z_2$ fluxes.
%The Trotterized dynamics and gate set-up for digital quantum simulator platforms are presented in Fig.~\ref{fig:digital_protocol}(d,e) of Appendix~B.

\emph{\color{tum}Concluding Remarks.--}
Quantum simulations of QSL properties should distinguish the realization of full-fledged topological order from the associated signatures of fractionalized edge excitations. Preparing topological order requires long-range entanglement in the ground state. By contrast, here we established that probing fractionalized chiral modes can be realized in quench dynamics from selected high-energy initial states. 
The key is to choose a short-range-entangled initial state that selects gauge-flux disorders compatible with chiral topology. The ensuing spin correlations average over this ensemble, dephasing nonuniversal oscillations and exposing the chiral signal.
The Cartesian product state we propose implements this principle by fixing triangular fluxes while leaving even-plaquette fluxes unresolved. The latter bind non-hybridizing Aharonov--Bohm-caged MZMs, preserving band topology of Majorana fermion excitations. 

%We emphasize that the exact Aharonov--Bohm caging relies on fine-tuned interference conditions \cite{Vidal98}. For example, digital gate dynamics breaks the compactness, e.g. first-order Trotterization generates next-nearest-neighbor Majorana hopping \cite{ali2025} violating compact localization \cite{supp}. As a result, zero modes hybridize which limits sharp recurrences of the edge signal to a transient observation window. However, we construct a zero-mode preserving circuit in Appendix~B which stroboscopically restores the caged MZM after every cycle. 

%We used the Yao-Kivelson model as a minimal example  of a more generic design principle. The local caging criterion [Eq.~\eqref{eq:cage}] and Cartesian product state [Eq.~\eqref{eq:state}] extend to all triangle-decorated trivalent models of Eq.~\eqref{eq:YK_spin_H}, without requiring translation symmetry. Our companion work shows how a fractal lattice enables similar protocols with versatile geometry-enabled features~\cite{companion_work}. Recently discovered amorphous and hyperbolic Kitaev liquids~\cite{CasselaOrnellasHodson2023,GrushinRepellin2023,Lenggenhager2024,Dusel2025} also motivate their decorated geometries in which to explore how geometry shapes the chiral dynamics. Such freedom allows the lattice to be tailored to a processor's qubit connectivity. The SM~\cite{supp} gives constructions on different qubit layouts of superconducting quantum processors with Aharonov--Bohm--caged MZMs; analogous fractionalized chiral dynamics is expected.

In the future, it will be interesting to study whether initial-state selection can also shape other dynamical protocols for observing or creating chiral states, for example in measurement-only set-ups for chiral phases~\cite{Sriram23,Ali23,Kai24} or for revealing Floquet topological order \cite{Levin13,Po2017,Will2025}. 
Recently discovered amorphous and hyperbolic Kitaev liquids~\cite{CasselaOrnellasHodson2023,GrushinRepellin2023,Lenggenhager2024,Dusel2025} also motivate their decorated geometries in which to explore how geometry shapes the chiral dynamics.
Whether topology leaves similarly sharp signatures in quench dynamics starting from high energy states beyond solvable spin-$\frac{1}{2}$ Kitaev-type models, for instance in generalized Heisenberg models with chiral QSL ground states, remains an interesting direction as well.

\acknowledgments{
{\it \color{tum}{Acknowledgments.-- }}
We are grateful to  P. d'Ornellas, F. Dusel, K. Kobayashi, Y. Lensky, P. Roushan, Y.-Z. You, F. Pollmann, and R. Pereira for discussions.   
S.F., M.K., and J.K. acknowledge support from the Deutsche Forschungsgemeinschaft (DFG, German Research Foundation) under Germany’s Excellence Strategy--EXC--2111--390814868, TRR 360 – 492547816 and DFG grants No. KN1254/1-2, KN1254/2-1, the Imperial-TUM flagship partnership, the European Union (grant agreement No. 101169765), as well as the Munich Quantum Valley, which is supported by the Bavarian state government with funds from the Hightech Agenda Bayern Plus. S.F. and M.K. were also supported in part by grant NSF PHY-2309135 to the Kavli Institute for Theoretical Physics (KITP) during the KITP Program: Engineering Fractionalization in Quantum Many-Body Systems: From New Quantum Materials to Quantum Simulators. W. N. acknowledges support from Conselho Nacional de Desenvolvimento Científico e Tecnológico (CNPq) for support through productivity grant 305739/2024-7 and the support of the INCT project Advanced Quantum Materials, involving the Brazilian agencies CNPq (Proc. 408766/2024-7), FAPESP (Proc. 2025/27091-3), and CAPES.
}

{\it \color{tum}{Data availability.-- }}
Data are available on Zenodo~\cite{zenodo}.

\bibliography{biblio}

\onecolumngrid
% \clearpage
\vspace{1em}
\begin{center}
    \textbf{End Matter}
\end{center}
\vspace{1em}
\twocolumngrid
{\it \color{tum} Appendix A: Aharonov--Bohm-caged MZMs.---}
We establish the generalized version of Eq.~\eqref{eq:cage} for any triangular-decorated lattice, with each even-sided plaquette surrounded by decorating triangles. We construct it by imposing zero energy and cancellation of outward propagation together at $J_\triangle/J_\parallel=1$. Consider a generic even-sided plaquette $R$ surrounded by triangles, with $n(p)=2q$ vertices labeled clockwise, such as a dodecagon in Fig.~\ref{fig:cmzm}(a). The pair $(2m-1,2m)$ belongs to a decorating triangle $\triangle^m$ whose third vertex $o_m$ lies outside $R$, for $m=1,\ldots,q$. With clockwise indices along the face, its flux is $W_{n(p)}=-\prod_{j=1}^{n(p)}u_{j,j+1}$.

We first establish the necessary condition. 
For a real wavefunction $\phi$ that vanishes outside $R$, the equation $A\phi=0$ reduces, after canceling the common factor $2J$, to
\begin{subequations}
 \label{eq:app_cage_local_equations}
 \begin{gather}
 u_{j,j+1}\phi_{j+1}=u_{j-1,j}\phi_{j-1},
 \label{eq:app_cage_on_face}\\[2pt]
 u_{o_m,2m-1}\phi_{2m-1}+u_{o_m,2m}\phi_{2m}=0.
 \label{eq:app_cage_no_leakage}
 \end{gather}
\end{subequations}
Equation~\eqref{eq:app_cage_on_face} imposes zero energy on the plaquette; Eq.~\eqref{eq:app_cage_no_leakage} cancels the two amplitudes reaching each exterior triangle tip. All remaining exterior equations vanish because those sites have no neighbors in $R$.
Equation.~\eqref{eq:app_cage_on_face} and~\eqref{eq:app_cage_no_leakage} give $|\phi_{j+1}|=|\phi_{j-1}|$ and $|\phi_{2m-1}|=|\phi_{2m}|$ respectively. Thus every site on $R$ has the same magnitude; normalization fixes $|\phi_j|=1/\sqrt{n(p)}$.

Multiplying Eq.~\eqref{eq:app_cage_on_face} by $\phi_j$ gives $u_{j,j+1}\phi_j\phi_{j+1}=u_{j-1,j}\phi_{j-1}\phi_j$, a common value associated with every bond; and we denote their common sign by $\tau\equiv n(p) u_{1,2}\phi_1\phi_2=\pm1$. Hence, on every bond, $u_{j,j+1}\phi_j\phi_{j+1}=\tau/n(p)=\tau\phi_j^2$. Dividing by the nonzero $\phi_j$ and using $u_{j,j+1}^2=1$ yields
\begin{equation}
 \phi_{j+1}=\tau u_{j,j+1}\phi_j.
 \label{eq:app_cage_recursion}
\end{equation}
For even $n(p)$, closure of Eq.~\eqref{eq:app_cage_recursion} gives $\prod_j u_{j,j+1}=1$ (clockwise), hence \mbox{$W_{n(p)}=-1$}, i.e. the even-plaquette flux must be flipped from the ground state. 
We now identify $\tau$ with the surrounding triangular flux. Consider the clockwise order on the adjacent triangle $(2m-1,o_m,2m)$, Eq.~\eqref{eq:app_cage_no_leakage} yields
\begin{equation}
 \begin{aligned}
 W_{n(\triangle^m)}
 &=-u_{2m-1,2m}u_{o_m,2m-1}u_{o_m,2m}\\
 &=u_{2m-1,2m}\frac{\phi_{2m}}{\phi_{2m-1}}=\tau,~~~\forall m =1,\cdots, q.
 \end{aligned}
 \label{eq:app_cage_triangle_flux}
\end{equation}
where for the last equality we used Eq.~\eqref{eq:app_cage_recursion} and $u_{j,j+1}^2=1$. 
Thus every surrounding triangle must have the same flux, establishing both requirements in Eq.~\eqref{eq:cage}.

Conversely, suppose these flux conditions ($W_{n(p)} = -1,\,W_3=\tau$)  hold. Set $\phi_1=1/\sqrt{n(p)}$ and construct the other amplitudes using Eq.~\eqref{eq:app_cage_recursion}. The condition $W_{n(p)}=-1$ ensures closure, while $u_{j,j+1}\phi_{j+1}=\tau\phi_j=u_{j-1,j}\phi_{j-1}$ satisfies Eq.~\eqref{eq:app_cage_on_face}. The triangle-flux product gives $\tau u_{2m-1,2m}=-u_{o_m,2m-1}/u_{o_m,2m}$, which makes the same recursion satisfy Eq.~\eqref{eq:app_cage_no_leakage}. Therefore it yields an exact caged MZM in $R$. Necessity and sufficiency of Eq.~\eqref{eq:cage} are thus established. 

Note that only the even-sided plaquette and its adjacent triangles enter this construction; other even-plaquette fluxes are unrestricted. For independent caged modes $\phi_a$ and $\phi_b$ of the same $A$, the matrix elements $\phi_a^T A\phi_b=0$ even when their supports overlap. They consequently remain at zero energy without mutual hybridization \cite{Tim26}, as used in the main text.

{\it \color{tum} Appendix B: Zero-Mode Preserving Trotterization.---}
Kitaev bond gates $\exp(\iu\theta\sigma_i^\alpha\sigma_j^\alpha)$ conserve the fluxes, but their Trotter errors can modify the matter hoppings and spoil caging~\cite{supp}. 
For example, first-order Trotterization applies all $z$, then $x$, then $y$ bond gates, each with angle $\theta = JT$ \cite{Will2025}. Its leading effective-Hamiltonian correction is of order $J^2T$ but includes next-nearest-neighbor Majorana hopping \cite{ali2025}. Symmetric second-order gates uses the five layers $x,y,z,y,x$, with angle $JT/2$ in the four outer layers and $JT$ in the central layer. This reduces the correction to $O(J^3T^2)$, but generally still violates the Aharonov–Bohm-caging condition.

We therefore need to design a zero-mode-preserving (ZMP) cycle that restores each original Aharonov--Bohm-caged MZM on an even-sided plaquette after every step. 
For exact preservation of the caged MZM $\phi$, separate the gates into intra- and inter-triangle bonds. Since the gates are gauge-invariant, we separate them in the matter fermion subspace: $A(u)=A_\triangle+A_\parallel$. Define the one-cycle digital Majorana propagator
\begin{equation}
 U_u(T)\equiv e^{t_\parallel A_\parallel/2}e^{t_\triangle A_\triangle}
 e^{t_\parallel A_\parallel/2}.
 \label{eq:app_digital_goal}
\end{equation}
Figures~\ref{fig:digital_protocol}(a1,a7) illustrate $ e^{t_\parallel A_\parallel/2}$ and Figs.~\ref{fig:digital_protocol}(a2-a6) illustrate $e^{t_\triangle A_\triangle}$. 
The parameters $t_\parallel(T)$ and $t_\triangle(T)$ are chosen to satisfy
\begin{equation}
    U_u(T)\phi=\phi
\end{equation}
for every Aharonov--Bohm-caged MZM on even plaquettes. 
The inter-triangle bonds connect disjoint pairs. Each two-site hopping matrix squares to $-4J^2$, hence $A_\parallel^2\phi=-4J^2\phi$, while zero energy gives $A_\parallel\phi=-A_\triangle\phi$. Parametrizing the inter-triangle evolution step by $T=\tan(Jt_\parallel)/J$ gives
\begin{equation}
 e^{\pm t_\parallel A_\parallel/2}\phi=
 \frac{(1\mp TA_\triangle/2)\phi}{\sqrt{1+(JT)^2}}.
 \label{eq:app_digital_bridge}
\end{equation}
Using Eq.~\eqref{eq:app_digital_goal} and multiplying $U_u(T)\phi=\phi$ on the left by $e^{-t_\parallel A_\parallel/2}$ reduces the restoration condition to
\begin{equation}
 e^{t_\triangle A_\triangle}(1-TA_\triangle/2)\phi
 =(1+TA_\triangle/2)\phi.
 \label{eq:app_digital_restore}
\end{equation}
We require this identity to hold for arbitrary $\phi$. An isolated equal-coupling triangle has eigenvalues $0,\pm2\iu\sqrt3J$. For eigenvalues, Eq.~\eqref{eq:app_digital_restore} becomes $e^{2\iu\sqrt3Jt_\triangle}=(1+\iu\sqrt3JT)/(1-\iu\sqrt3JT)$. This fixes $t_\parallel=\arctan(JT)/J$ and $t_\triangle=\arctan(\sqrt3JT)/(\sqrt3J)$. $e^{t_\triangle A_\triangle}$ thus maps $e^{t_\parallel A_\parallel/2}\phi$ to $e^{-t_\parallel A_\parallel/2}\phi$, and the final layer $e^{t_\parallel A_\parallel/2}$ in Fig.~\ref{fig:digital_protocol}(a7) restores $\phi$. 

Moreover, the triangle evolution $e^{t_\triangle A_\triangle}$ is thus resolved into two-qubit gates exactly by the rotations in Fig.~\ref{fig:digital_protocol}(a2-a6):
\begin{align}
 \theta_x&=\tfrac12\arctan(JT), \label{eq:app_digital_angles1}\\
 \theta_y&=\tfrac12\arctan\!\frac{JT}{\sqrt{1+(JT)^2}}, \label{eq:app_digital_angles2}\\
 \theta_z&=\arctan\!\frac{JT}{\sqrt{1+2(JT)^2}}.
 \label{eq:app_digital_angles3}
\end{align}
for the steps (2-6) in Fig.~\ref{fig:digital_protocol}(a). Details can be found in SM~\cite{supp}.
The total seven-layer cycle for Fig.~\ref{fig:digital_protocol}(a) has a second order error in $O(J^3T^2)$, the same order as symmetric Trotterization, while its action on the caged MZMs remains identity.

Figure~\ref{fig:digital_protocol} compares both on-site quench spin dynamics and OTOC after $n$ cycles. At small evolution step $JT=0.01$, all three protocols resolve chiral recurrences and repeated directional OTOCs; the results using ZMP and the other two Trotter gates nearly coincide. At $JT=0.5$, while a larger Trotter error changes the chiral velocity thus the winding period, the early-time insets still show a clear first OTOC arrival for every protocol within about ten cycles. The first-order sequence can therefore be sufficient for short-time measurements. However, at longer times, the ZMP cycle retains much stronger recurrences and OTOC recurrences than either conventional Trotterization. 
The advantage of the ZMP protocol is thus most pronounced when sustained recurrences are needed at a larger usable $JT$ and longer observation times. We point out that the circuit presented in Fig.~\ref{fig:digital_protocol}(a) is not the unique solution to ZMP protocol. %Other constructions and efficient implementation on quantum processors are promising open questions for future research.  

\end{document}

\begin{center}
    {\large\bfseries Supplemental Material for ``\ourtitle''\par}
    \medskip
    Shi Feng\,\orcidlink{0000-0002-0184-7049}$^{1,2}$,
    Willian M. H. Natori\,\orcidlink{0000-0002-0740-2956}$^{3}$,
    Michael Knap\,\orcidlink{0000-0002-7093-9502}$^{1,2}$,
    and Johannes Knolle\,\orcidlink{0000-0002-0956-2419}$^{1,2,4}$\par
    \smallskip
    {\small\itshape
    $^1$Technical University of Munich, TUM School of Natural Sciences, Physics Department, 85748 Garching, Germany\par
    $^2$Munich Center for Quantum Science and Technology (MCQST), Schellingstr. 4, 80799 M{\"u}nchen, Germany\par
    $^3$Gleb Wataghin Institute of Physics, University of Campinas, Campinas, S\~ao Paulo 13083-950, Brazil\par
    $^4$Blackett Laboratory, Imperial College London, London SW7 2AZ, United Kingdom\par}
\end{center}
% Shared numbering and contents for standalone and attached SM.
\setcounter{section}{0}
\setcounter{subsection}{0}
\setcounter{subsubsection}{0}
\setcounter{figure}{0}
\setcounter{table}{0}
\setcounter{equation}{0}
\renewcommand{\thesection}{S\arabic{section}}
\renewcommand{\thefigure}{S\arabic{figure}}
\renewcommand{\thetable}{S\arabic{table}}
\renewcommand{\theequation}{S\arabic{equation}}
\setcounter{secnumdepth}{3}
\setcounter{tocdepth}{2}
\tableofcontents

\section{Yao-Kivelson Chiral Spin Liquid} \label{subsec:YK_model_A}
The model is defined on the star lattice, obtained by replacing every site of the honeycomb lattice with a triangle \cite{YaoKivelson2007}. The resulting lattice is trivalent: each site is connected to one $x$ bond, one $y$ bond, and one $z$ bond. 
The two elementary plaquette types have different roles: the triangular fluxes select a time-reversal-breaking background, whereas flipped dodecagons support the Aharonov-Bohm cage presented in the main text.

The spin Hamiltonian is the Kitaev exchange model
\begin{equation}
    H = -\sum_{\langle ij\rangle_\alpha}
    J_{\langle ij\rangle_\alpha} \,
    \sigma_i^\alpha \sigma_j^\alpha ,
    \;\;
    \alpha=x,y,z,
    \label{eq:sm_YK_spin_H}
\end{equation}
where $\langle ij\rangle_\alpha$ denotes a nearest-neighbor bond of type
$\alpha$.  In general, triangle bonds and inter-triangle bonds connecting neighboring triangles may have different couplings.  For example, one may write
\begin{equation}
    J_{\langle ij\rangle_\alpha} =
    \begin{cases}
        J_\alpha^{\triangle}, & 
        \text{ inside a decorating triangle},\\
        J_\alpha^{\parallel}, & 
        \text{ between decorating triangles}.
    \end{cases}
    \label{eq:YK_bond_couplings}
\end{equation}
Unless otherwise stated, all bonds have the same isotropic Pauli-spin coupling $J>0$. We set $\hbar=1$, quote energies in $J$, and use the dimensionless time $Jt$.

\begin{figure}[t]
    \centering
    \includegraphics[width=0.5\linewidth]{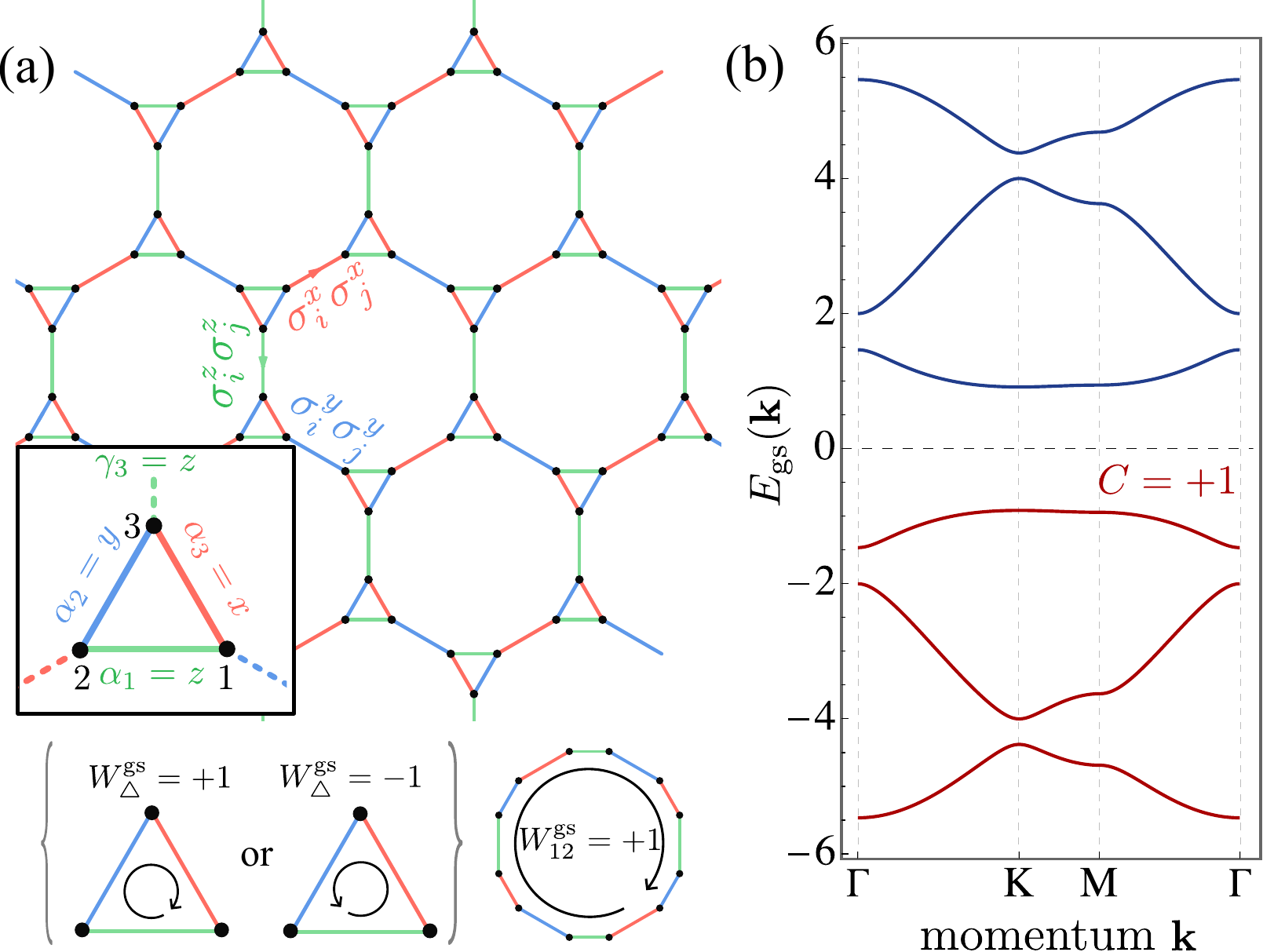}
    \caption{
    \textbf{Ground-state sector of the star-lattice Kitaev model.}
    (a) Top: Star-lattice geometry obtained by decorating the honeycomb lattice with
    triangles.  The red, blue, and green bonds denote the Kitaev
    $x$, $y$, and $z$ spin exchanges, respectively; and fluxes denote the ground-state flux sector. The inset illustrates the definition of Eq.~\eqref{eq:plaquette_flux_spin} for $W_\triangle = -\iu^{3}\prod_{j=1}^{3} \bigl(\sigma_j^{\alpha_j}\sigma_{j+1}^{\alpha_j}\bigr) = \sigma_1^{\gamma_1} \sigma_2^{\gamma_2} \sigma_3^{\gamma_3}$.  
    Bottom: ground-state flux pattern.  The dodecagon plaquettes carry
    $W_{12}^{\rm g.s.}=+1$, while the decorating triangles select a
    definite chiral sector with $W_\triangle=+1$ or its time-reversed partner $W_\triangle=-1$. 
    (b) Gapped Majorana bands in the ground-state flux sector along the high-symmetry path $\Gamma$-$K$-$M$-$\Gamma$.  The occupied
    negative-energy bands (in red) carry Chern number $C=+1$, giving a flux-trapped MZM and a chiral Majorana edge mode on an open boundary.
    }
    \label{fig:star_lattice}
\end{figure}

The Hamiltonian has an extensive set of local conserved fluxes. Number the vertices of a plaquette $p$ clockwise as
$1,\ldots,n_p$, with $n_p+1\equiv1$. Let $\alpha_j$ be the color of bond $(j,j+1)$ and $\gamma_j$ the color leaving the plaquette at site $j$, see the inset of Fig.~\ref{fig:star_lattice}(a) for example. The conserved flux in the spin basis reads:
\begin{equation}
    W_p=-\iu^{\,n_p}\prod_{j=1}^{n_p} \bigl(\sigma_j^{\alpha_j}\sigma_{j+1}^{\alpha_j}\bigr) = \prod_{j=1}^{n_p}\sigma_j^{\gamma_j}.
    \label{eq:plaquette_flux_spin}
\end{equation}
The bond factors are ordered from $j=1$ on the left to $j=n_p$ on the right. These operators satisfy
$[W_p,H]=0,\;\;[W_p,W_{p'}]=0,\;\;W_p^2=1$. A many-body eigenstate can therefore be labeled by a configuration of
$\mathbb Z_2$ plaquette fluxes.  On the star lattice, we denote the triangle fluxes by $W_\triangle$ and the dodecagon fluxes by $W_{12}$.  The key preparation distinction in this work is that $W_\triangle$ will be fixed, while $W_{12}$ will be left unresolved.

The exact solution represents each spin by four Majorana operators, $\sigma_i^\alpha=\iu b_i^\alpha c_i$, with the physical-state constraint $D_i=b_i^x b_i^y b_i^z c_i=1$. On an oriented $\alpha$ bond, define the link operator
\begin{equation}
 \hat u_{ij}\equiv\iu b_i^\alpha b_j^\alpha,
 \;\;\hat u_{ji}=-\hat u_{ij},\;\;\hat u_{ij}^{\,2}=1.
 \label{eq:link_operator}
\end{equation}
In the enlarged Majorana Hilbert space,
$[\hat u_{ij},H]=[\hat u_{ij},\hat u_{kl}]=0$.
These conserved link operators are the emergent $\mathbb{Z}_2$ gauge field whose physical sectors are to be fixed by the physical fluxes $W_p$. Before fixing a sector, the Hamiltonian is quartic: $H=(\iu/4)\sum_{ij}\hat A_{ij}c_ic_j$, with $\hat A_{ij}=2J_{\langle ij\rangle_\alpha}\hat u_{ij}$ on a bond, $\hat A_{ji}=-\hat A_{ij}$. 
In a fixed gauge sector $u=\{u_{ij}=\pm1\}$, replace each conserved link operator $\hat u_{ij}$ by its eigenvalue $u_{ij}$. The Hamiltonian then reduces to a Majorana tight-binding model
\begin{equation}
 H(u)=\frac{\iu}{4}\sum_{ij}A_{ij}(u)c_ic_j,
 \label{eq:majorana_H}
\end{equation}
where the real antisymmetric hopping matrix has entries $A_{ij}(u)=2J_{\langle ij\rangle_\alpha}u_{ij}$ on an $\alpha$ bond.

The spin and link descriptions of the flux are directly connected by
$\sigma_i^\alpha\sigma_j^\alpha=-\iu\hat u_{ij}c_ic_j$.
In the ordered loop product, adjacent $c$ fermions cancel via
$(c_1c_2)(c_2c_3)\cdots(c_{n_p}c_1)=1$.
Substitution in Eq.~\eqref{eq:plaquette_flux_spin} therefore gives, on physical states,
\begin{equation}
 W_p=-\prod_{j=1}^{n_p} \hat u_{j,j+1}
 \;\;{\rm for~}n_p=3,12.
 \label{eq:flux_sign_conversion}
\end{equation}
We therefore use $W_p =-\prod_{j=1}^{n_p} u_{j,j+1}$ for its eigenvalue when the gauge sector is understood.
Note that time reversal reverses the three-spin product $W_\triangle$, whereas $W_{12}$ remains even. 

For a fixed flux sector, the Majorana spectrum comes in particle-hole symmetric pairs $\operatorname{spec}\bigl(iA(u)\bigr)=\{+\epsilon_n(u),-\epsilon_n(u)\},\;\epsilon_n\ge 0$.
The fermionic vacuum energy in sector $u$ is $E_{\rm vac}(u)=-\frac{1}{2}\sum_{\epsilon_n>0}\epsilon_n(u)$. 
The usual ground-state problem would require choosing the flux sector minimizing the vacuum energy and then occupying the corresponding negative Majorana levels.

Yao and Kivelson identified the energy-minimizing uniform flux sector through perturbative and numerical analysis~\cite{YaoKivelson2007}.
In the spin convention of Eq.~\eqref{eq:plaquette_flux_spin}, the two time-reversed ground-state patterns are
\begin{equation}
 W_\triangle^{\rm gs}=\pm1,\;\;
 W_{12}^{\rm gs}=+1,
 \label{eq:even_flux_reference}
\end{equation}
The dodecagon values $W_{12}^{\rm gs} = +1$ correspond to a $\pi$ flux in the fermion basis, consistent with Lieb's canonical even-loop rule \cite{Lieb1994}.
Selecting either sign for the triangular flux spontaneously breaks time reversal, giving the two gapped chiral spin-liquid ground-state sectors with Chern number $|C| = 1$ near isotropic coupling $J^{\triangle}_\alpha/ J_\alpha^{\parallel} = 1$, as illustrated in Fig.~\ref{fig:star_lattice}(b).

\section{Disorder-Averaged Majorana Spectral Functions}
\label{subsec:spectral_function}
The cylinder in Fig.~2 of the main text is open in the transverse $L_x$ direction and periodic around the $L_y$ circumference. Label its periodic cells by $m=0,\ldots,L_y-1$ and combine the transverse cell and six sublattice indices into $\lambda=1,\ldots,6L_x$.
For each sampled link field $u$, we diagonalize the full $6L_xL_y$-site matrix $h(u)=\iu A(u)$, with eigenvalues $\varepsilon_n^{(u)}$ and normalized wavefunctions $\varphi_n^{(u)}(m,\lambda)$, in a fixed gauge convention.
The broadened real-space spectral kernel is
\begin{equation}
 \mathcal A_u(m,\lambda; m',\lambda';\omega)
 =\langle m,\lambda|\delta_\eta[\omega-h(u)]|m',\lambda'\rangle
 =\sum_n\varphi_n^{(u)}(m,\lambda)
          \varphi_n^{(u)*}(m',\lambda')
          \delta_\eta(\omega-\varepsilon_n^{(u)}),
 \label{eq:real_space_spectral_kernel}
\end{equation}
with $\delta_\eta(x)=\eta/[\pi(x^2+\eta^2)]$ the Lorentzian broadening factor.

To define the spatial average, first choose a relative separation $\rho=(m-m')\bmod L_y$, with $\rho=0,\ldots,L_y-1$, and write the two cell coordinates as $(m,m')=(R+\rho,R)\bmod L_y$. Here $R=m'$ is the reference cell.
For each fixed $\rho$, summing over $R=0,\ldots,L_y-1$ translates the pair together around the tube without changing its separation. 
On an unwrapped copy of the circumference, the midpoint is
$(m+m')/2=R+\rho/2$. At fixed $\rho$, averaging over $R$ is therefore equivalent to averaging over this center-of-mass position \cite{Feng2025}. 

We perform this spatial average at fixed $u$, then Fourier transform the separation $\rho$ and trace over all transverse orbitals:
\begin{equation}
 A(k_\parallel,\omega;u)
 =\sum_{\rho=0}^{L_y-1}e^{-\iu k_\parallel\rho}
       \sum_{\lambda=1}^{6L_x}
    \left[\frac{1}{L_y}\sum_{R=0}^{L_y-1}
 \mathcal A_u(R+\rho,\lambda;R,\lambda;\omega)\right].
 \label{eq:average_momentum_spectrum}
\end{equation}
All cell indices are modulo $L_y$, and
$k_\parallel=2\pi\ell/L_y\pmod{2\pi}$, with
$\ell=0,\ldots,L_y-1$.  The bracketed spatial average is taken within a single realization. 
To evaluate this quantity, define the longitudinal Fourier component
\begin{equation}
 \widetilde\varphi_n^{(u)}(k_\parallel,\lambda)
 =\frac{1}{\sqrt{L_y}}\sum_{m=0}^{L_y-1}
 e^{-\iu k_\parallel m}\varphi_n^{(u)}(m,\lambda).
 \label{eq:longitudinal_fourier_component}
\end{equation}
Substitute Eq.~\eqref{eq:real_space_spectral_kernel} into
Eq.~\eqref{eq:average_momentum_spectrum} and change variables from
$(R,\rho)$ to $(m,m')=(R+\rho,R)\bmod L_y$. Hence
\begin{equation}
 \begin{aligned}
 A(k_\parallel,\omega;u)
 =\sum_n\sum_{\lambda=1}^{6L_x}
 \left|\widetilde\varphi_n^{(u)}(k_\parallel,\lambda)\right|^2
\delta_\eta(\omega-\varepsilon_n^{(u)}).
 \end{aligned}
\end{equation}
The spectra are then averaged over independent disorder realizations:
\begin{equation}
 A(k_\parallel,\omega)=\frac{1}{N_s}\sum_{a=1}^{N_s}
                   A(k_\parallel,\omega;u_a),
\end{equation}
which is used for obtaining the average-momentum resolved Majorana spectra in Fig.~2 of the main text.

\begin{figure}[t]
    \centering
    \includegraphics[width=0.5\linewidth]{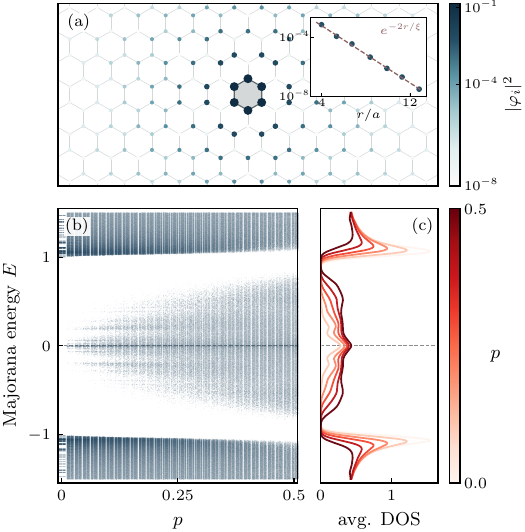}
    \caption{\textbf{Non-compact flux-bound MZM and flux-induced gap filling in the honeycomb model.} (a) Center of a lattice with open boundary conditions with one flipped hexagon, shaded gray. Color and dot size show the normalized MZM intensity $|\varphi_i|^2$ on a logarithmic color scale. Inset: shell-averaged intensity versus distance $r/a$, where $a$ is the NN bond length. (b) Majorana energy levels versus hexagonal flux-flip probability $p$ on a $30\times30$ torus, with 20 realizations per probability. (c) DOS of the same ensemble. Red shades indicate $p$. All panels use an NNN-to-NN hopping ratio of $0.2$.}
    \label{fig:honeycomb_flux_disorder_gap}
\end{figure}

\section{Contrast: Disorder-Induced MZM Hybridization in Honeycomb Lattice} \label{sec:honey}
The honeycomb model provides a comparison in which flux-bound Majorana modes have a finite localization length. Consider the isotropic Kitaev model supplemented by chiral NNN Majorana hopping. Its flux-free sector is gapped and has $|C|=1$, while a flipped hexagon binds a vortex Majorana mode~\cite{Kitaev2006,Lahtinen2011}. A spin state with unresolved hexagonal fluxes therefore samples Hamiltonians containing different configurations of these defects through the sector average.

Figure~\ref{fig:honeycomb_flux_disorder_gap}(a) shows the spatial structure of one such mode. A single hexagon is flipped on a large open patch, with its companion Majorana near the outer boundary. Within the lowest two-dimensional particle--hole subspace, we choose the normalized real combination that maximizes the weight near the vortex core. This separates the core profile from its boundary partner without imposing a spatial cutoff. The logarithmic intensity map and its radial inset reveal weak tails extending beyond the central hexagon, with an approximately exponential envelope $e^{-2r/\xi}$. The finite $\xi$ permits overlap between modes bound to distinct vortices.

When several vortices occupy the same sample, their isolated-core solutions generally mix and split away from zero energy. The splitting decreases exponentially with separation and can oscillate in sign~\cite{Lahtinen2011}. At finite density, these couplings form a low-energy Majorana hopping network whose structure depends on the vortex arrangement. Random vortex backgrounds can consequently support a class-D thermal metal~\cite{Senthil2000,Chalker2001,Huse2012}; the individual defect modes have no geometric constraint enforcing exact zero coupling throughout the ensemble.

The spectra in Fig.~\ref{fig:honeycomb_flux_disorder_gap}(b) show how increasing the hexagonal flux-flip probability $p$ fills the clean gap with levels distributed around zero. The corresponding averaged DOS in panel (c) develops a broad low-energy contribution \cite{Senthil2000,Chalker2001}, in contrast to the levels pinned at $E=0$ under even-only star-lattice disorder [Fig.~2(f) of the main text].

\begin{figure}[t]
\centering
\includegraphics[width=0.5\linewidth]{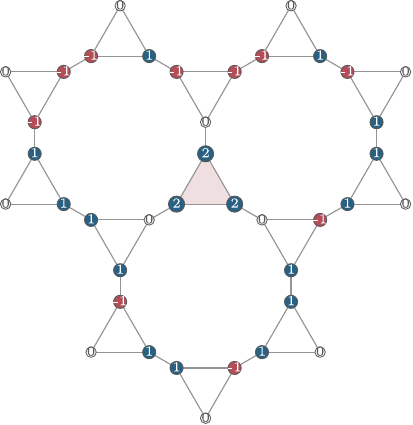}
\caption{\textbf{A large compact triangle-defect mode in a specified gauge.} The shaded central triangle is flipped; its three neighboring dodecagons retain their reference flux. Numbers give the integer amplitudes $6\phi_j$ on the 30-site candidate patch; all other sites have zero amplitude. Blue and red distinguish positive and negative values.}
\label{fig:triangle_wf}
\end{figure}

\section{Zero Modes for Triangular Fluxes} \label{app:trimzm}
A triangle flux defect is not covered by the even-plaquette Aharonov-Bohm caging criterion, and a naive construction of an associated triangular MZM cannot satisfy $A \phi = 0$. 
We test a larger compact MZM support $S$ associated with a triangular flux flip.
For one reversed interior triangle in an otherwise uniform-flux star lattice, the solution $S$ is the union of its three adjacent dodecagons. It contains $30$ distinct sites, as is shown in Fig.~\ref{fig:triangle_wf}:
24 sites have amplitudes $\pm1/6$, three central sites have $2/6$, and
three sites of the candidate patch have zero amplitude. 
Further nearby triangle flips change the entries of $A$ and can
remove this zero-energy solution. Note this solution, although compact, spans a much larger area, distinct from the even-plaquette flux flip which traps an exact Aharonov-Bohm cage on the plaquette with zero localization length. This indicates why the topological invariant and the edge mode do not survive an all-plaquette disorder ensemble.

\section{Topological Invariant}
\label{sec:topological_invariant}
We compute the real-space Chern marker, which determines whether the unresolved flux sectors sampled by the initial state share the same quantized Majorana Chern number despite the absence of translation symmetry. 

For each fixed flux configuration $u$, the spin Hamiltonian reduces to a quadratic Majorana problem with Hermitian single-particle Hamiltonian $h(u)=\iu A(u)$ and particle-hole-symmetric spectrum.  Because a random flux configuration breaks translation symmetry, momentum-space Berry curvature is
not available sector by sector.  We instead use the real-space Chern marker of Bianco and Resta~\cite{Bianco11}, which is defined directly from the single-particle spectral projector and applies naturally to disordered finite samples, and has been widely used in determining the topological nature in amorphous or disordered systems \cite{Adhip23}. We note that other types of real-space Chern marker also exist \cite{peru22}, which produces the same bulk topological invariant.

Let $P$ denote the projector onto the occupied Majorana subspace and $Q=1-P$ its complement.  With $X=\sum_i x_i|i\rangle\langle i|$ and $Y=\sum_i y_i|i\rangle\langle i|$, the local marker at site $i$ is 
\begin{equation}
    C_i
    \sim
    -4\pi\,{\rm Im}\,
    \langle i|
        P X Q Y P
    |i\rangle .
    \label{eq:chern_marker}
\end{equation}
up to normalization.
A technical point arises from the exact Majorana zero modes bound to flipped even plaquettes.  Denoting the negative energy and zero-energy projectors by
\begin{equation}
    P_-
    =
    \sum_{\varepsilon_n<0}
    |\varphi_n\rangle\langle\varphi_n|,
    \;\;
    P_0
    =
    \sum_{\varepsilon_n=0}
    |\varphi_n\rangle\langle\varphi_n|,
    \label{eq:negative_zero_projectors}
\end{equation}
the spectral projector at half filling is ambiguous exactly at
$\varepsilon=0$.  We therefore use either of the two canonical choices
\begin{equation}
    P=P_-
    \;\;\text{or}\;\; P=P_-+P_0 ,
    \label{eq:zero_mode_projector_choices}
\end{equation}
corresponding to placing the reference energy infinitesimally below or above
the zero-mode manifold.  Including only an arbitrary subset of the degenerate zero modes would not be basis independent, since numerical diagonalization may return arbitrary, apparently delocalized linear combinations within the same compact subspace.  By contrast, the full projector $P_0$ is uniquely defined.
Because this subspace consists of compact localized modes and carries no extended states, including or excluding it does not change the local Chern marker.

\begin{figure*}[t]
    \centering
    \includegraphics[width=0.9\linewidth]{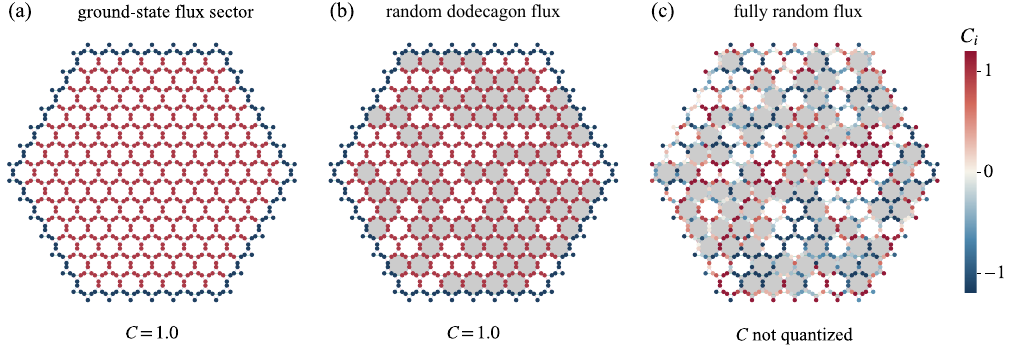}
    \caption{
    \textbf{Local Chern marker of the Yao-Kivelson model under flux disorder.} Gray plaquettes indicate fluxes flipped relative to the ground-state sector.
    (a) In the ground-state flux sector, the bulk local marker is quantized at $C=1$.
    (b) Randomizing only dodecagon plaquettes leaves the bulk Chern marker quantized, showing that the Aharonov-Bohm cage with $\xi = 0$ does not destroy the topology of the Majorana sector.
    (c) Randomizing all plaquettes destroys the stable quantized Chern number, giving $C \simeq 0$ on average.
    }
    \label{fig:chern_markers}
\end{figure*}

For a finite lattice with open boundaries, the sum of $C_i$ over the entire sample vanishes.  The nonzero bulk contribution is compensated by an oppositely signed boundary contribution, visible as the blue outer region in Fig.~\ref{fig:chern_markers}(a) and (b).  The bulk invariant must therefore be extracted from a region $\mathcal R_{\rm bulk}$ away from the physical edge:
\begin{equation}
    C =\frac{1}{|\mathcal R_{\rm bulk}|}\sum_{i\in\mathcal R_{\rm bulk}} C_i
    \label{eq:bulk_chern_marker}
\end{equation}
Figure~\ref{fig:chern_markers}(a) shows the ground-state flux sector for the chosen triangle chirality.  Indeed, the marker rapidly approaches $C_i=1$ away from the open edge and remains approximately uniform throughout the interior, and Eq.~\eqref{eq:bulk_chern_marker} gives $C=1.0$ up to small corrections.  

The central result relevant for the chiral dynamics is shown in Fig.~\ref{fig:chern_markers}(b).  Randomizing the  dodecagon plaquette fluxes leaves the bulk marker pinned to the same quantized value, despite the extensive Aharonov-Bohm-cage manifold.  The Aharonov-Bohm cages fill the spectral gap at precisely zero energy, but they do not provide extended
states that connect the occupied and unoccupied continua.  The mobility gap of the propagating Majorana sector, and hence its Chern invariant, remains intact. Thus every flux configuration with random fluxes on dodecagons conditioned by the Cartesian product state belongs to the same chiral phase.

This behavior sharply contrasts with fully random fluxes in Fig.~\ref{fig:chern_markers}(c).  Once the triangle fluxes are also randomized, the local marker develops a spatially irregular mixture of positive and negative values, with no quantized bulk plateau.  The corresponding $C$ is not pinned to an integer and is not stable as a bulk invariant.
This loss of quantization is consistent with the hybridized low-energy continuum discussed in Sec.~\ref{app:trimzm}: triangle-flux defects destroy the uniform chiral bias and close the mobility gap.

%%%%%%%%%%%%%%%%%%%%%%%%%%%
%%%%%%%%%%%%%%%%%%%%%%%%%%%

\section{Next-Nearest-Neighbor Hopping and Loss of Compactness}
\label{sec:sm_nnn_perturbation}
\begin{figure}[!ht]
 \centering
 \includegraphics[width=.5\linewidth]{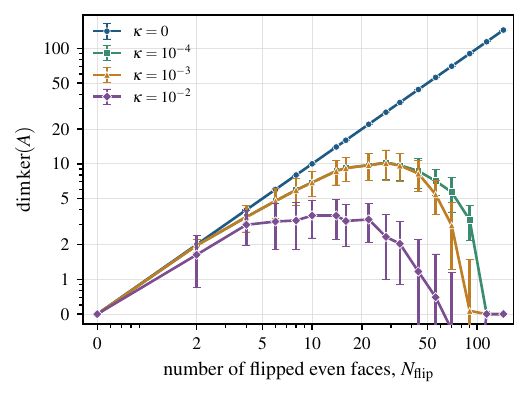}
 \caption{\textbf{Lifting of the compact zero-mode degeneracy by NNN hopping.} An exact number $N_{\rm flip}$ of even plaquettes is flipped on a periodic star lattice, with all triangle fluxes held in the same chiral sector. The plotted zero-mode count is averaged over random fluxes, with error bars indicating variation between configurations. At $\kappa=0$, the compatible configurations obey $\dim[\ker A_0]=N_{\rm flip}$. The finite-$\kappa$ curves count eigenvalues within a fixed numerical tolerance of $10^{-6}$.}
 \label{fig:sm_nnn_kernel}
\end{figure}

The compact-mode protection relies on the cancellation of every hopping channel leaving the supporting face. Conserving the plaquette fluxes alone does not enforce this condition. To examine its failure, we add the flux-conserving length-two hopping used in Fig.~\ref{fig:sm_nnn_kernel}, with strength controlled by $\kappa$. A path $i\to j\to k$ contributes a matter bilinear proportional to $i u_{ij}u_{jk}c_ic_k$, with an orientation-dependent sign. Its physical spin representation is a three-spin Pauli exchange. Paths crossing an inter-triangle bond introduce genuine next-nearest-neighbor (NNN) hopping, while paths lying within a decorating triangle modify an existing hopping. 

Write the perturbed Majorana Hamiltonian as $A_\kappa(u)=A_0(u)+\delta A_\kappa(u)$. An unperturbed Aharonov-Bohm cage $\phi_a$, supported on a flipped even plaquette $R_a$, satisfies $A_0\phi_a=0$ and vanishes outside $R_a$. Its response to the added hopping is therefore
\begin{equation}
 A_\kappa\phi_a=\delta A_\kappa\phi_a,\;\;
 (A_\kappa\phi_a)_j=\sum_{k\in R_a}(\delta A_\kappa)_{jk}\phi_{a,k}
 \;\; (j\notin R_a).
 \label{eq:sm_nnn_leakage}
\end{equation}
The new outward channels generally give a nonzero right-hand side, bypassing the nearest-neighbor cancellation. The original compact vector consequently leaves the exact zero-energy subspace. In an orthonormal basis of the unperturbed zero-mode space, the antisymmetric matrix $M_{ab}=\phi_a^T\delta A_\kappa\phi_b$ gives the leading defect-mode couplings. Nonzero couplings split pairs of Majorana modes away from zero and allow their wavefunctions to extend beyond the original plaquette.

Figure~\ref{fig:sm_nnn_kernel} contrasts the extensive exact zero-mode subspace at $\kappa=0$ with its suppression under the added NNN hopping perturbation. The finite-$\kappa$ count depends on the numerical tolerance, which we set to $10^{-6}$. %Together, the departure from the nearest-neighbor reference and Eq.~\eqref{eq:sm_nnn_leakage} show how flux conservation can coexist with the loss of exact Aharonov-Bohm cages.
At finite defect density, hybridization generates a low-energy Majorana hopping network, which can support a disorder-induced class-D metal~\cite{Huse2012,wang2024}. Whether the mobility gap closes depends on the localization length of these states; in-gap localized modes due to sparse defects may coexist with a quantized Chern number; while a gapless continuum due to flux proliferation will trivialize the band topology.

\begin{figure}[b]
 \centering
 \includegraphics[width=.75\linewidth]{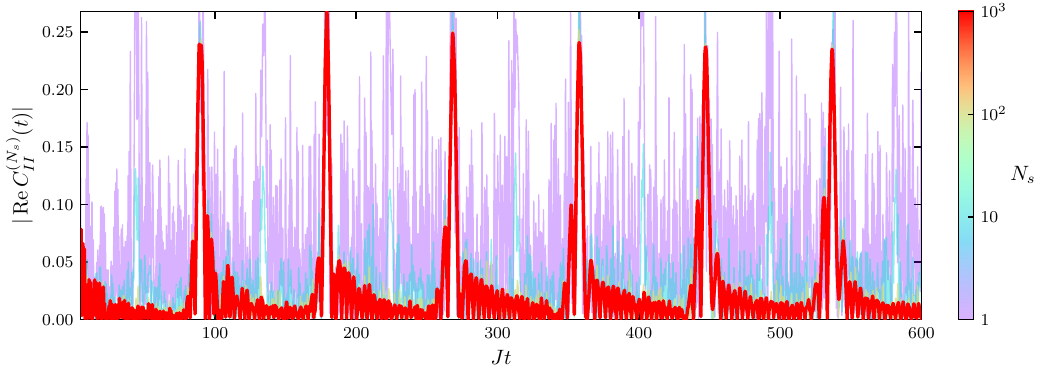}
 \caption{\textbf{Cartesian-state on-site correlations $|\operatorname{Re}C_{II}^{(N_s)}(t)|$ on a 288-site lattice, through $Jt=600$}. The rainbow colorbar gives the number of sampled even-flux sectors; the thicker red curve uses $N_s=1000$. Return amplitudes are averaged before taking the absolute value. }
 \label{fig:sm_dephasing}
\end{figure}

\section{Dynamical Dephasing}
\label{sec:sm_dephasing}
For the Cartesian state, sampling independent even-flux sectors from their prepared-state weights gives the finite-ensemble estimator
\begin{equation}
 C_{II}^{(N_s)}(t)=\frac{1}{N_s}\sum_{a=1}^{N_s}
                         [e^{tA(u_a)}]_{II}.
 \label{eq:sm_sampled_return}
\end{equation}
The return is real because $A(u)$ is real and antisymmetric. Figure~\ref{fig:sm_dephasing} plots $|\operatorname{Re}C_{II}^{(N_s)}(t)|$ for increasing flux-disorder ensembles. Averaging is performed before taking the absolute value.  With few sectors (purple), non-equilibrium oscillations obscure the recurrent structure. Increasing the number of samples, $N_s$, suppresses the flux-sector-dependent background, and the recurrence approaches a sequence of well-resolved pulses (red).

\section{Zero-Mode-Preserving Digital Evolution}
\label{sec:sm_cmzm}
\subsection{Exact zero-mode restoration}
Separate triangle and inter-triangle exchanges as $H_0=H_\triangle+H_{\parallel}$, with $A_0=A_\triangle+A_{\parallel}$ and $A_{0,ij}=2Ju_{ij}$. For a real compact vector, $A_0\phi=0$ gives $A_\triangle\phi=-A_{\parallel}\phi$.
Begin with two unknown evolution parameter $t_{\parallel}$ and $t_\triangle$; $T$ will label the resulting calibrated family. The cycle, its first inter-triangle gate, and the restoration condition are
\begin{subequations}
 \begin{align}
 U_u(T)&=D_u e^{t_\triangle A_\triangle}D_u,\;\;
 D_u=e^{t_{\parallel}A_{\parallel}/2},
 \label{eq:sm_grouped_cycle}\\
 U_u(T)\phi=\phi
 &\;\Longleftrightarrow\;
 e^{t_\triangle A_\triangle}D_u\phi=D_u^{-1}\phi.
 \label{eq:sm_restoration_condition}
 \end{align}
\end{subequations}
Equation~\eqref{eq:sm_restoration_condition} follows by multiplying the full-cycle equation on the left by $D_u^{-1}$. The middle pulse must map $D_u\phi$ to $D_u^{-1}\phi$, which the final forward inter-triangle pulse restores to $\phi$. Inter-triangle bonds are disjoint pairs. On a matched pair their block is $2J\begin{pmatrix}0&u_{ij}\\-u_{ij}&0\end{pmatrix}$, whose square is $-4J^2$. Unmatched boundary sites carry zero weight in these modes. In the inter-triangle exponential, the relation
\begin{equation}
    A_{\parallel}^2\phi=-4J^2\phi
\end{equation}
makes every even power proportional to $\phi$ and every odd power to $A_{\parallel}\phi$:
\begin{equation}
 \begin{aligned}
 e^{\pm t_{\parallel}A_{\parallel}/2}\phi
 &=\sum_{m=0}^{\infty}\frac{(-1)^m(Jt_{\parallel})^{2m}}{(2m)!}\phi
 \;\;\pm\frac{A_{\parallel}}{2J}
 \sum_{m=0}^{\infty}\frac{(-1)^m(Jt_{\parallel})^{2m+1}}{(2m+1)!}\phi\\
 &=\cos(Jt_{\parallel})\phi
       \pm\frac{\sin(Jt_{\parallel})}{2J}A_{\parallel}\phi.
 \end{aligned}
 \label{eq:sm_bridge_action}
\end{equation}
The first sum is the cosine series and the second is the sine series. Insert the $+$ and $-$ versions of Eq.~\eqref{eq:sm_bridge_action} for $D_u\phi$ and $D_u^{-1}\phi$, respectively, into Eq.~\eqref{eq:sm_restoration_condition}. Using $A_{\parallel}\phi=-A_\triangle\phi$ and dividing by the nonzero $\cos(Jt_{\parallel})$ on the branch near zero pulse area gives
\begin{equation}
 e^{t_\triangle A_\triangle}
 \left(1-\frac{\tan(Jt_{\parallel})}{2J}A_\triangle\right)\phi
 =\left(1+\frac{\tan(Jt_{\parallel})}{2J}A_\triangle\right)\phi.
 \label{eq:sm_triangle_goal}
\end{equation}
To satisfy Eq.~\eqref{eq:sm_triangle_goal} for every compact vector with one calibration, we impose the stronger, sufficient matrix identity
\[
 e^{t_\triangle A_\triangle}
 \left(1-\frac{\tan(Jt_{\parallel})}{2J}A_\triangle\right)
 =1+\frac{\tan(Jt_{\parallel})}{2J}A_\triangle.
\]
Because $A_\triangle$ is block diagonal, we solve the identity on the three matter-site amplitudes of each decorating triangle. The compact mode obeys $A_0\phi=0$ for the combined generator: the generally nonzero triangle and inter-triangle contributions cancel through $A_\triangle\phi=-A_{\parallel}\phi$. The eigenvalues needed to design the triangle evolution are,
\begin{equation}
 \left.A_\triangle\right|_{\{1,2,3\}}
 =2J\begin{pmatrix}0&u_{12}&-u_{31}\\-u_{12}&0&u_{23}\\u_{31}&-u_{23}&0\end{pmatrix},\;\;
 A_\triangle^3=-12J^2A_\triangle.
 \label{eq:sm_triangle_cube}
\end{equation}
The prefactor $2J$ follows from the Pauli-exchange convention. In a fixed link sector, one exchange $H_{ij}=-J\sigma_i^\alpha\sigma_j^\alpha=iJu_{ij}c_ic_j$ gives $\dot c_i=i[H_{ij},c_i]=2Ju_{ij}c_j$, hence $A_{ij}=2Ju_{ij}$. For the displayed triangle block, expanding the characteristic determinant gives
\[
 \begin{aligned}
 \det\!\left[\lambda 1_3-\left.A_\triangle\right|_{\{1,2,3\}}\right]
 &=\lambda\left[\lambda^2+4J^2(u_{12}^2+u_{23}^2+u_{31}^2)\right]\\
 &=\lambda(\lambda^2+12J^2).
 \end{aligned}
\]
where we used $u_{ij}^2=1$. Setting this determinant to zero yields $\lambda=0,\pm2i\sqrt3J$. The Hermitian single-particle matrix $iA_\triangle$ thus has real eigenvalues $0,\pm2\sqrt3J$. Direct multiplication gives $(A_\triangle^3)_{12}=-24J^3u_{12}$.

Both sides of the triangular evolution matrix identity are functions of $A_\triangle$ and share its eigenvectors. On an eigenvector with eigenvalue $\lambda$, each matrix function therefore becomes the corresponding scalar function of $\lambda$. The required equality is
\[
 e^{t_\triangle\lambda}
 \left(1-\frac{\tan(Jt_{\parallel})}{2J}\lambda\right)
 =1+\frac{\tan(Jt_{\parallel})}{2J}\lambda.
\]
For $\lambda=0$ both sides equal one. The two nonzero eigenvalues give complex-conjugate equations for real evolution parameters, so it suffices to solve the $\lambda=+2i\sqrt3J$ case. Substituting this eigenvalue gives $\lambda\tan(Jt_{\parallel})/(2J)=i\sqrt3\tan(Jt_{\parallel})$, and hence
\begin{equation}
 \begin{aligned}
 e^{2i\sqrt3Jt_\triangle}\left[1-i\sqrt3\tan(Jt_{\parallel})\right]
 &=1+i\sqrt3\tan(Jt_{\parallel}),\\
 e^{2i\sqrt3Jt_\triangle}
 &=\frac{1+i\sqrt3\tan(Jt_{\parallel})}{1-i\sqrt3\tan(Jt_{\parallel})}
 =e^{2i\arctan[\sqrt3\tan(Jt_{\parallel})]},\\
 \tan(\sqrt3Jt_\triangle)&=\sqrt3\tan(Jt_{\parallel}).
 \end{aligned}
 \label{eq:sm_triangle_phase}
\end{equation}
The ratio on the second line has unit modulus and the displayed phase. Equating phases modulo $2\pi$ gives the final tangent relation. 
For Hamiltonian simulation we label this family by $T\equiv\tan(Jt_{\parallel})/J$. For a specified $JT$, we have
\begin{equation}
 t_{\parallel}=\frac{\arctan(JT)}{J},\;\;
 t_\triangle=\frac{\arctan(\sqrt3JT)}{\sqrt3J}.
 \label{eq:sm_calibrated_durations}
\end{equation}
The equation $\tan(Jt_{\parallel})=JT$ is an exact parametrization of the restoring family. Together with the symmetric ordering, it makes $T$ the simulated Hamiltonian step to second order.
The full triangle rotation thus obeys
\begin{equation}
    e^{t_\triangle A_\triangle}(1-TA_\triangle/2)=1+TA_\triangle/2.
\end{equation}
Using $A_{\parallel}\phi=-A_\triangle\phi$, Eq.~\eqref{eq:sm_bridge_action} gives $D_u\phi=\cos(Jt_{\parallel})[1-\tan(Jt_{\parallel})A_\triangle/(2J)]\phi$. Equation~\eqref{eq:sm_calibrated_durations} sets $\tan(Jt_{\parallel})=JT$ and $\cos(Jt_{\parallel})=1/\sqrt{1+(JT)^2}$, yielding the first vector in the complete sequence:
\begin{equation}
 D_u\phi=\frac{(1-TA_\triangle/2)\phi}{\sqrt{1+(JT)^2}}
 \;\xrightarrow{\;e^{t_\triangle A_\triangle}\;}\;
 \frac{(1+TA_\triangle/2)\phi}{\sqrt{1+(JT)^2}}=D_u^{-1}\phi
 \;\xrightarrow{\;D_u\;}\;\phi.
 \label{eq:sm_mode_restoration}
\end{equation}

\subsection{Two-qubit gates for \texorpdfstring{$\exp( A_\triangle t_\triangle)$}{the triangle evolution}}
The grouped triangle pulse still has to be built from two-qubit gates. For this calculation, number its vertices so that $P_x=\sigma_1^x\sigma_2^x$, $P_y=\sigma_2^y\sigma_3^y$, and $P_z=\sigma_3^z\sigma_1^z$. Each operator acts on two spins and squares to one. Each pair shares one site with different Pauli axes; for example, $P_xP_y=i\sigma_1^x\sigma_2^z\sigma_3^y=-P_yP_x$. They therefore anticommute pairwise, giving $(P_x+P_y+P_z)^2=3$. Summing the even and odd powers in the exponential then gives $e^{iJt_\triangle(P_x+P_y+P_z)}=\cos(\sqrt3Jt_\triangle)\,1+i\sin(\sqrt3Jt_\triangle)(P_x+P_y+P_z)/\sqrt3$. Equation~\eqref{eq:sm_calibrated_durations} sets $\sqrt3Jt_\triangle=\arctan(\sqrt3JT)$, so $\cos(\sqrt3Jt_\triangle)=1/\sqrt{1+3(JT)^2}$ and $\sin(\sqrt3Jt_\triangle)/\sqrt3=JT/\sqrt{1+3(JT)^2}$. Thus the desired spin unitary is exactly
\begin{equation}
 e^{iJt_\triangle(P_x+P_y+P_z)}
     =\frac{1+i(JT)(P_x+P_y+P_z)}{\sqrt{1+3(JT)^2}}.
 \label{eq:sm_triangle_spin_target}
\end{equation}
We now solve for unknown angles in the symmetric ansatz $G_x(\theta_x)G_y(\theta_y)G_z(\theta_z)G_y(\theta_y)G_x(\theta_x)$, where $G_\alpha(\theta)=e^{i\theta P_\alpha}$. We evaluate this product from its center outwards, keeping the gate order fixed. In the following expressions, $G_\alpha$ means $G_\alpha(\theta_\alpha)$. Since $P_\alpha^2=1$,
\[
 G_\alpha=\cos\theta_\alpha\,1+i\sin\theta_\alpha P_\alpha,
 \;\;
 G_\alpha^2=\cos(2\theta_\alpha)\,1+i\sin(2\theta_\alpha)P_\alpha.
\]
For $\alpha\neq\beta$, the equal-sign product appearing in the symmetric five-gate sequence has the expansion
\[
 \begin{aligned}
 G_\alpha P_\beta G_\alpha
 &=(\cos\theta_\alpha\,1+i\sin\theta_\alpha P_\alpha)
 P_\beta(\cos\theta_\alpha\,1+i\sin\theta_\alpha P_\alpha)\\
 &=\cos^2\theta_\alpha P_\beta
 +i\cos\theta_\alpha\sin\theta_\alpha(P_\alpha P_\beta+P_\beta P_\alpha)
 -\sin^2\theta_\alpha P_\alpha P_\beta P_\alpha\\
 &=(\cos^2\theta_\alpha+\sin^2\theta_\alpha)P_\beta=P_\beta.
 \end{aligned}
\]
The mixed term vanishes by anticommutation, and $P_\alpha P_\beta P_\alpha=-P_\beta$ supplies the last sign. 
Now expand the central $G_z$ and apply the identity with $\alpha=y$, $\beta=z$:
\[
 \begin{aligned}
 G_yG_zG_y
 &=G_y(\cos\theta_z\,1+i\sin\theta_zP_z)G_y
 =\cos\theta_z G_y^2+i\sin\theta_z G_yP_zG_y\\
 &=\cos\theta_z\cos(2\theta_y)\,1
   +i\cos\theta_z\sin(2\theta_y)P_y+i\sin\theta_zP_z.
 \end{aligned}
\]
Next apply the two outer $G_x$ factors. Their action on the identity gives $G_x^2$, while $G_xP_yG_x=P_y$ and $G_xP_zG_x=P_z$ give
\begin{equation}
 \begin{aligned}
 G_xG_yG_zG_yG_x
 &=\cos\theta_z\cos(2\theta_y)G_x^2
   +i\cos\theta_z\sin(2\theta_y)G_xP_yG_x+i\sin\theta_zG_xP_zG_x\\
 &=\cos\theta_z\cos(2\theta_y)\cos(2\theta_x)\,1
 +i\cos\theta_z\cos(2\theta_y)\sin(2\theta_x)P_x
 +i\cos\theta_z\sin(2\theta_y)P_y+i\sin\theta_zP_z.
 \end{aligned}
 \label{eq:sm_unknown_angle_product}
\end{equation}
Equating the four coefficients to Eq.~\eqref{eq:sm_triangle_spin_target} determines the angles. The ratio of the $P_x$ and identity coefficients gives $\tan(2\theta_x)=JT$. Adding their squares gives $\cos\theta_z\cos(2\theta_y)=\sqrt{1+(JT)^2}/\sqrt{1+3(JT)^2}$, and division into the $P_y$ coefficient gives $\tan(2\theta_y)=JT/\sqrt{1+(JT)^2}$. Finally the sum of the first three squared coefficients fixes $\cos\theta_z=\sqrt{1+2(JT)^2}/\sqrt{1+3(JT)^2}$, which determines the $P_z$ coefficient. Hence we obtain
\begin{equation}
 \theta_x=\frac12\arctan(JT),\;\;
 \theta_y=\frac12\arctan\frac{JT}{\sqrt{1+(JT)^2}},\;\;
 \theta_z=\arctan\frac{JT}{\sqrt{1+2(JT)^2}}.
 \label{eq:sm_gate_angles}
\end{equation}

\begin{figure}[b]
 \centering
 \includegraphics[width=0.9\linewidth]{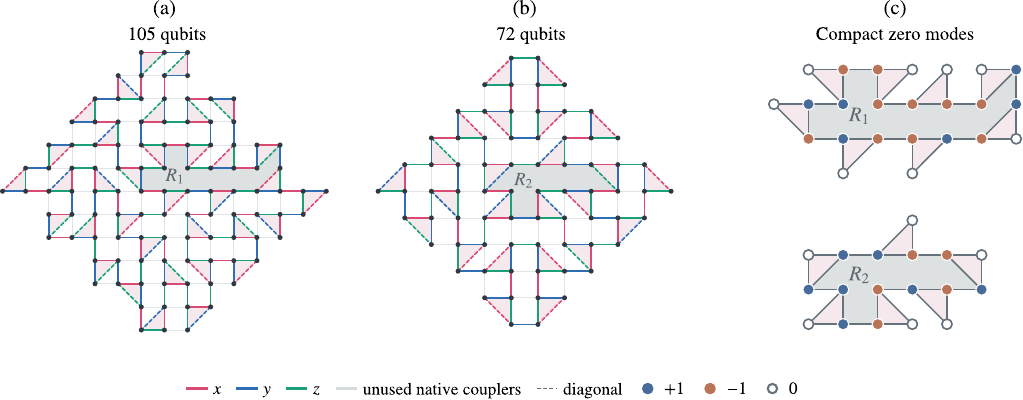}
 \caption{\textbf{Triangle decoration on superconducting qubit quantum processor layouts}. (a) All 105  qubits form 35 triangles and 11 complete even plaquettes. (b) All 72  qubits form 24 triangles and 9 complete even plaquettes; the layout is rotated so that its longer dimension is horizontal. Red, blue, and green denote $x,y,z$ exchanges. Light pink identifies decorating triangles, and faint gray segments are unused native couplers. Solid colored bonds use native couplers; dashed colored diagonals denote exchanges compiled from native gates. (c) Two compact modes: the upper example is supported on the gray plaquette $R_1$ in (a), and the lower on $R_2$ in (b). Filled circles show relative amplitudes $\pm1$, normalized by $1/4$ and $1/\sqrt{12}$, respectively; open circles have zero amplitude. Each selected plaquette has clockwise loop product $+1$, and its surrounding triangles have product $-1$. Every omitted amplitude is zero.}
 \label{fig:sm_google_geometry}
\end{figure}

\section{Triangle-Decorated Kitaev Model on Superconducting Quantum Processors}
\label{sec:sm_google_geometry}
Trapped-ion processors with all-to-all pairwise gate connectivity offer one route to implementing the geometry required \cite{Moses2023,ali2025}. The local compact-mode construction can also be realized on platforms with fixed, sparse connectivity since the geometry-induced destructive interference is local. Figure~\ref{fig:sm_google_geometry}(a,b) gives explicit triangle decorations of superconducting qubit quantum processors with 105-qubit and 72-qubit  layouts~\cite{Acharya2025,Google72}. Every qubit belongs to a complete decorating triangle, using all 105 and 72 sites. The resulting connected graphs contain 35 and 24 triangles, respectively, with properly colored bonds and complete even plaquettes of varying shape and perimeter. The irregular chip boundaries require no removal of qubits. These constructions extend the design beyond a Yao-Kivelson star lattice while retaining the local connectivity needed for compactness.

The two compact solutions in Fig.~\ref{fig:sm_google_geometry}(c) correspond to individually flipping the gray plaquettes $R_1$ and $R_2$ in panels (a) and (b), respectively, while keeping a common chirality on all triangles. Their relative amplitudes are $\pm1$ on the plaquette vertices and zero everywhere else; normalization gives magnitudes $1/4$ on the sixteen-site support and $1/\sqrt{12}$ on the twelve-site  support. The two Majorana hopping amplitudes meeting at each open exterior site cancel, and the support amplitudes satisfy the internal zero-energy equations. The same construction supplies one independent caged MZM per flipped even plaquette on both layouts, including configurations with several neighboring flips. Plaquette shape and translation symmetry do not enter this cancellation; the required ingredients are the triangle decoration, compatible fluxes, and equal hopping magnitudes.

The Cartesian state preparation chooses the missing intra-triangle color at each site and local spin orientations that fix a common triangle flux value. With these known, conserved fluxes, the calibrated triangle gate can be compiled exactly into three Pauli-exchange rotations on its two native edges. No additional ancilla qubit or even-flux readout is required. Separating the inter-triangle gates from inner triangle gates gives the same compact-mode restoration as Sec.~\ref{sec:sm_cmzm}. These layouts therefore accommodate the preparation and evolution protocol using native digital gates. Their boundary length and gate schedule can be chosen to suit a device.

\bibliography{biblio}